\documentclass[%
reprint,
superscriptaddress,
 amsmath,amssymb,
 aps,
prd,
]{revtex4-2}

\usepackage{graphicx}% Include figure files
\usepackage{dcolumn}% Align table columns on decimal point
\usepackage{bm}% bold math
\begin{document}

%\preprint{APS/123-QED}

\title{Alternative LISA-TAIJI networks}

% Force line breaks with \\
%\thanks{A footnote to the article title}%

\author{Gang Wang}
\email[Gang Wang: ]{gwang@shao.ac.cn, gwanggw@gmail.com}
%\homepage[]{Your web page}
%\thanks{}
\affiliation{Shanghai Astronomical Observatory, Chinese Academy of Sciences, Shanghai 200030, China}

\author{Wei-Tou Ni}
\email[Wei-Tou Ni: ]{weitou@gmail.com}
\affiliation{State Key Laboratory of Magnetic Resonance and Atomic and Molecular Physics, Innovation Academy for Precision Measurement Science and Technology (APM), Chinese Academy of Sciences, Wuhan 430071, China}
\affiliation{Department of Physics, National Tsing Hua University, Hsinchu, Taiwan, 30013, ROC}

\author{Wen-Biao Han}
\email[Wen-Biao Han: ]{wbhan@shao.ac.cn}
\affiliation{Shanghai Astronomical Observatory, Chinese Academy of Sciences, Shanghai 200030, China}
\affiliation{Hangzhou Institute for Advanced Study, University of Chinese Academy of Sciences, Hangzhou 310124, China}
\affiliation{School of Astronomy and Space Science, University of Chinese Academy of Sciences, Beijing 100049, China}

\author{Peng Xu}
\email[Peng Xu: ]{xp@lzu.edu.cn}
\affiliation{Institute of Mechanics, Chinese Academy of Sciences, Beijing 100190, China}
\affiliation{Lanzhou Center for Theoretical Physics, Lanzhou University, Lanzhou 730000, China}
\affiliation{Hangzhou Institute for Advanced Study, University of Chinese Academy of Sciences, Hangzhou 310124, China}

\author{Ziren Luo}
\email[Ziren Luo: ]{luoziren@imech.ac.cn}
\affiliation{Hangzhou Institute for Advanced Study, University of Chinese Academy of Sciences, Hangzhou 310124, China}
\affiliation{Institute of Mechanics, Chinese Academy of Sciences, Beijing 100190, China}
\affiliation{Taiji Laboratory for Gravitational Wave Universe (Beijing/Hangzhou), University of Chinese Academy of Sciences, Beijing 100049, China}

\date{\today}% It is always \today, today,

\begin{abstract}

The space-borne gravitational wave (GW) detectors, LISA and TAIJI, are planned to be launched in the 2030s. The dual detectors with comparable sensitivities will form a network observing GW with significant advantages. 
In this work, we investigate the three possible LISA-TAIJI networks for the different location and orientation compositions of LISA orbit ($+60^\circ$ inclination and trailing the Earth by $20^\circ$) and alternative TAIJI orbit configurations including TAIJIp ($+60^\circ$ inclination and leading the Earth by $20^\circ$), TAIJIc ($+60^\circ$ inclination and co-located with LISA), TAIJIm ($-60^\circ$ inclination and leading the Earth by $20^\circ$). In the three LISA-TAIJI configurations, the LISA-TAIJIm network shows the best performance on the sky localization and polarization determination for the massive binary system due to their better complementary antenna pattern, and LISA-TAIJIc could achieve the best cross-correlation and observe the stochastic GW background with an optimal sensitivity. 

\end{abstract}

\maketitle

\section{Introduction}

The gravitational wave detection, GW150914, was observed by Advanced LIGO detectors at two sites Hanford, WA and Livingston, LA \cite{Abbott:2016blz}. Two interferometers are designed to be (closely) aligned interferometric arms with a separation of 3000 km. The GW170814 and GW170817 were the first detections coincidently observed by triple interferometers of Advanced LIGO and Advanced Virgo. As benefits of the misaligned orientation between LIGO and Virgo detectors, the source directions were well localized and the alternative GW polarizations were tested \cite{Abbott:2017oio,GW170817:detection,GW170817:TGR}. The KAGRA detector is expected to join the ground-based interferometer network in near future \cite{KAGRA:2020,LVK:LLR}. The detector network surrounding the Earth will improve the angular resolution of the sky localization and parameter determination on the GW sources \citep{LVK:LLR,Schutz:2011tw}.
Although the current GW detections are all from the compact binary coalescences, Advanced LIGO and Advanced Virgo are actively searching for the stochastic GW background (SGWB) \cite{TheLIGOScientific:2016dpb,Abbott:2018utx,LIGOScientific:2019vic,Abbott:2021xxi,Abbott:2021ksc,Abbott:2021jel}. 
The detection of the stochastic relic GW will deeply impact our understanding on the early Universe. To distinguish cosmological imprint from the instrument noise and astrophysical foreground, joint observations from two or more independent detectors are highly demanded. 

The multiple interferometer cooperation is also planned in the next-generation space missions for GW observation in the deci-Hz middle frequency band. Both the BBO and DECIGO missions proposed three constellations deploying on the Earth-like heliocentric orbit with $120^\circ$ separations \cite{BBO:2005,DECIGO:2006}. The number of detectors will increase the signal-to-noise ratio (SNR) of the detection, and the large separations between the constellations will improve their angular resolution of the sky localization for the sources. The SGWB is also expected to be observed by the two co-located and coplanar interferometers of the BBO or DECIGO \cite{Romano:2016dpx,Schmitz:2020syl}. The present activities of GW missions in this middle frequency band are briefly reviewed in \cite{Ni:2020utm}.

The space-borne missions targeting for the milli-Hz low frequency band GW observation including LISA \cite{2017arXiv170200786A}, TAIJI \cite{Hu:2017mde}, and TianQin \cite{Luo:2015ght} are scheduled to be launched around the 2030s. Each of the missions will include a triangular constellation formed by three spacecraft (S/C).  The LISA and TAIJI missions are designed to be the heliocentric orbit. To achieve stable interferometer arms, the S/C formation plane of the LISA/TAIJI is designed to be $\pm 60^\circ$ with the ecliptic by employing the Clohessy–Wiltshire framework \cite{Dhurandhar:2004rv}. By assuming the TAIJI is leading the Earth by $20^\circ$ and LISA is trailing the Earth by $20^\circ$, \citet{Ruan:2020smc} and \citet{Wang:2020a} investigated the sky localization improvement of the LISA-TAIJI network compared to the single LISA mission. \citet{Omiya:2020fvw,Seto:2020mfd}, and \citet{Orlando:2020oko} evaluated network capabilities for the SGWB observation. \citet{Wang:2020dkc} and \citet{Wang:2021srv} estimated the impact of the joint LISA-TAIJI observation on cosmological parameter determination. And \citet{Wang:2021polar} demonstrated the observation constraints on the GW polarizations from the joint observation.

Considering the orbital configuration of the TAIJI mission is not fully determined, the merits of the alternative LISA-TAIJI network are worth evaluating. In this work, by presetting the LISA orbit is determined, we investigate the performances of three possible LISA-TAIJI networks for different TAIJI orbital selections as shown in Fig. \ref{fig:LISA_TAIJI}, a) TAIJIp which leading the Earth by $\sim$20$^\circ$ and the formation of the constellation is $+60^\circ$ inclined as LISA, b) TAIJIm which also leading the Earth by $\sim$20$^\circ$ and the plane of the S/C has a $-60^\circ$ inclination compared to the LISA, and c) TAIJIc which is co-located and coplanar with LISA and trailing the Earth by $\sim$20$^\circ$. 
The deployment and observation for the TAIJI mission from these orbit choices are expected to be not too much different. However, the joint observations with LISA from alternative TAIJI mission orbits could yield different performances for the supermassive black hole (SMBH) binary and SGWB observations. We evaluate networks' angular resolutions of sky localization for the SMBH binary, observations for the alternative polarizations beyond general relativity (GR), and the overlap reduction function for the SGWB observations. In the three pairs combination, the LISA-TAIJIm network demonstrates the best parameter determinations for the SMBH binary system, and LISA-TAIJIc shows an optimal capability for the SGWB observation. 

This paper is organized as follows. 
In Sec. \ref{sec:alternative_LISA_TAIJI}, we introduce three LISA-TAIJI network configurations and their joint sensitivities. 
In Sec. \ref{sec:SMBH_results}, we report and compare the results of parameter determinations on the SMBH binary from three LISA-TAIJI networks including the angular resolution and the alternative polarization constraints.
In Sec. \ref{sec:stochastic_background}, we investigate the overlap reduction functions of the three networks for the SGWB observation.
We recapitulate our conclusions in Sec. \ref{sec:conclusions}. (We set $G=c=1$ in this work except otherwise stated).

\section{Alternative LISA-TAIJI networks} \label{sec:alternative_LISA_TAIJI}

\subsection{The LISA and TAIJI orbital configurations}

\begin{figure*}[htb]
\includegraphics[width=0.48\textwidth]{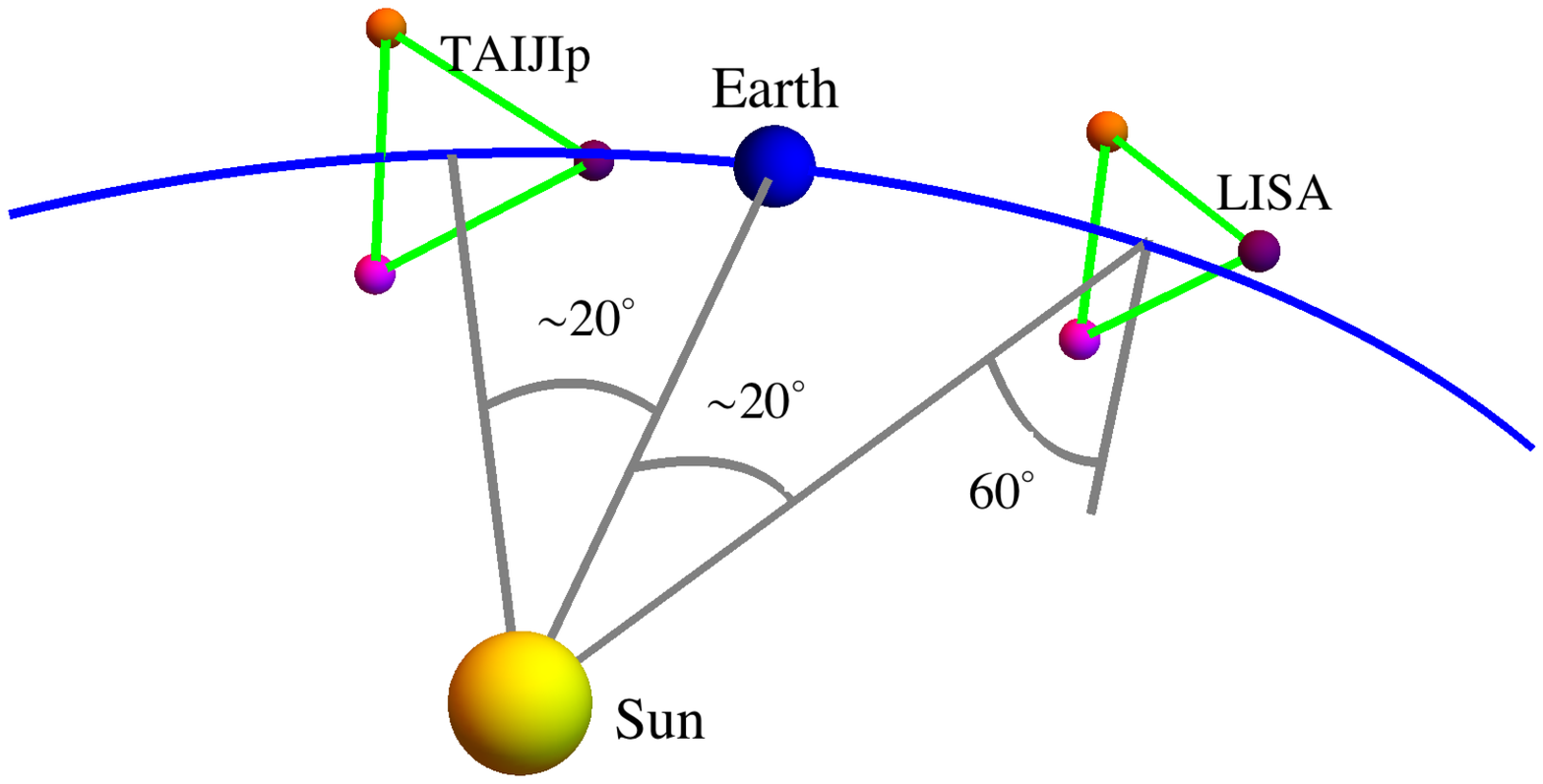}
\includegraphics[width=0.48\textwidth]{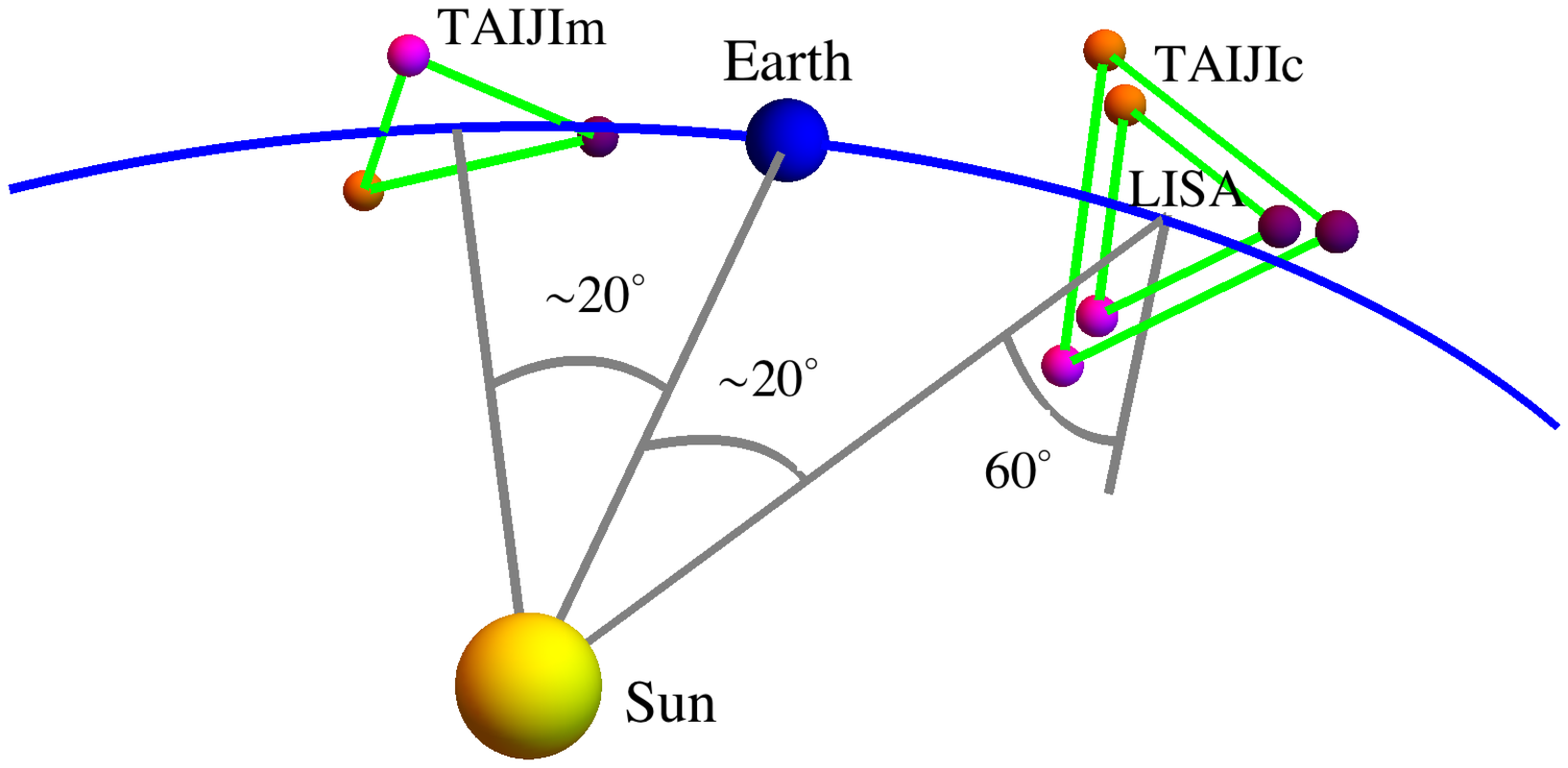}
\caption{\label{fig:LISA_TAIJI} The diagram of LISA and TAIJI mission orbital configurations. The left panel shows the LISA (trailing the Earth by $\sim$20$^\circ$ and $+60^\circ$ inclined with respect to the ecliptic plane) and the TAIJIp (leading the Earth by $\sim$20$^\circ$ and $+60^\circ$ inclined as the LISA). The right panel shows the LISA and anther two optional TAIJI orbital choices which are the TAIJIm (leading the Earth by $\sim$20$^\circ$ and with a $-60^\circ$ inclination) and TAIJIc (coplanar and co-located with LISA). The angle between the LISA and TAIJIp formation planes is $\sim$34.5$^\circ$, and the angle between the LISA and TAIJIm formation planes is $\sim$71$^\circ$.
}
\end{figure*}

The LISA mission is scheduled to be launched in the 2030s which includes three S/C forming a $2.5 \times 10^6$ km triangle trailing the Earth by $20^{\circ}$ \cite{2017arXiv170200786A}.
The constellation plane has a 60$^{\circ}$ inclination with respect to the ecliptic plane as shown in Fig. \ref{fig:LISA_TAIJI}. The TAIJI mission is proposed as a LISA-like orbital configuration with a $3 \times 10^6$ km arm length \cite{Hu:2017mde}. An assumed orbit for the TAIJI is that the constellation is in front of the Earth by 20$^{\circ}$ and has the same $60^\circ$ inclination as LISA as shown in the left plot of Fig. \ref{fig:LISA_TAIJI}, and this TAIJI orbital configuration is labeled as TAIJIp in this work. The preset $20^\circ$ trailing/leading angle is a practical compromise from the launch vehicle, telemetry capabilities, and the gravitational perturbation reduction \cite{LISA2000}.

The TAIJI orbital configuration could also have other choices without (significantly) increasing the launch budget. The first alternative is that the constellation formation is tuned to be $-60^\circ$ inclination compared to the TAIJIp's $+60^\circ$ and we label it as TAIJIm. Another case would be that the TAIJI is co-located and coplanar with the LISA which is named as TAIJIc in this work. These two orbital configurations are shown in the right panel of Fig. \ref{fig:LISA_TAIJI}. 
For the LISA-TAIJIc network, their orientations of S/C formation would be co-aligned. For the LISA-TAIJIp, the angle between the two formation planes is $\sim$34.5$^\circ$, their separation angle will be $\sim$40$^\circ$ and distance is $\sim1 \times 10^8$ km. And the angle of the orientations between the LISA and TAIJIm is around $ 71^\circ$.

In this work, by employing the numerical orbits, we investigate the performances of three pairs of LISA-TAIJI networks (LISA-TAIJIp, LISA-TAIJIm, and LISA-TAIJIc) on the detectability for SMBH binaries and the SGWB.
These three combinations should cover all the possible dual detector scenarios except the LISA with a co-located TAIJI having a $-60^\circ$ inclination which would be insipid because neither the large separation benefiting for the compact binary observation nor strong cross-correlation for the SGWB observation is expected. 
The numerical orbits for the TAIJIp and TAIJIc are from our work in \cite{Wang:2017aqq,Wang:2020a}, and the orbit for TAIJIm is newly obtained from our optimization method in \cite{Wang:2012aea,Wang:2012ce,Wang:2013te,Dhurandhar:2011ik,Wang:2014cla,Wang:2017aqq}. 

\subsection{Response formulation of TDI channel} \label{secsub:TDI_response}

For the space-borne GW missions, time-delay interferometry (TDI) is essential to suppress the laser frequency noise and achieve targeting sensitivity. The sensitivities for the different TDI channels have been evaluated numerically in our recent works \cite{Wang:1stTDI,Wang:2ndTDI}. With implementing the TDI, the GW response is combined from the response of each evolved single link.
The response functions to the GW tensor polarizations from GR in Doppler measurement were formulated in \cite{1975GReGr...6..439E,1987GReGr..19.1101W,Vallisneri:2007xa,Vallisneri:2012np}. And the response functions for the polarizations beyond the GR were developed in \citet{Tinto:2010hz}. To keep the integrity of the work, we reiterate the formulas of the response of TDI to the six polarizations as utilized in \cite{Wang:2021polar}.

The GW propagation vector from a source locating at ecliptic longitude $\lambda$ and latitude $\theta$ (in the solar-system barycentric coordinates) will be
\begin{equation} \label{eq:source_vec}
 \hat{k}  = -( \cos \lambda \cos \theta, \sin \lambda \cos \theta ,  \sin \theta ).
\end{equation} 
The polarization tensors of the GW signal for the $+$, $\times$, scalar breathing (b), scalar longitudinal (L), vector x and y, combining with the factors of the source's inclination angle $\iota$ are
\begin{widetext}
\begin{equation} \label{eq:polarizations-response}
\begin{aligned}
{\rm e}_{+} & \equiv \mathcal{O}_1 \cdot 
\begin{pmatrix}
1 & 0 & 0 \\
0 & -1 & 0 \\
0 & 0 & 0
\end{pmatrix}
\cdot \mathcal{O}^T_1 \times \frac{1+\cos^2 \iota}{2} ,
\qquad
{\rm e}_{\times}  \equiv \mathcal{O}_1 \cdot 
\begin{pmatrix}
0 & 1 & 0\\
1 & 0 & 0 \\
0 & 0 & 0
\end{pmatrix}
\cdot \mathcal{O}^T_1 \times i (- \cos \iota ), 
\\
{\rm e}_\mathrm{b} & \equiv \mathcal{O}_1 \cdot 
\begin{pmatrix}
1 & 0 & 0\\
0 & 1 & 0 \\
0 & 0 & 0
\end{pmatrix}
\cdot \mathcal{O}^T_1 \times \sin^2 \iota,  
\qquad \qquad \quad
{\rm e}_\mathrm{L}  \equiv \mathcal{O}_1 \cdot 
\begin{pmatrix}
0 & 0 & 0\\
0 & 0 & 0 \\
0 & 0 & 1
\end{pmatrix}
\cdot \mathcal{O}^T_1 \times \sin^2 \iota,
\\
{\rm e}_\mathrm{x} & \equiv \mathcal{O}_1 \cdot 
\begin{pmatrix}
0 & 0 & 1\\
0 & 0 & 0 \\
1 & 0 & 0
\end{pmatrix}
\cdot \mathcal{O}^T_1 \times \sin \iota \cos \iota, 
\qquad \quad \ 
{\rm e}_\mathrm{y}  \equiv \mathcal{O}_1 \cdot 
\begin{pmatrix}
0 & 0 & 0\\
0 & 0 & 1 \\
0 & 1 & 0
\end{pmatrix}
\cdot \mathcal{O}^T_1  \times i \sin \iota, 
\end{aligned}
\end{equation}
with
\begin{equation}
\mathcal{O}_1 =
\begin{pmatrix}
\sin \lambda \cos \psi - \cos \lambda \sin \theta \sin \psi & -\sin \lambda \sin \psi - \cos \lambda \sin \theta \cos \psi & -\cos \lambda \cos \theta  \\
     -\cos \lambda \cos \psi - \sin \lambda \sin \theta \sin \psi & \cos \lambda \sin \psi - \sin \lambda \sin \theta \cos \psi & -\sin \lambda \cos \theta  \\
         \cos \theta \sin \psi & \cos \theta \cos \psi & -\sin \theta 
\end{pmatrix},
\end{equation}
where $\psi$ is the polarization angle. The response to the GW polarization p in the link from S/C$i$ to $j$ will be
\begin{equation} \label{eq:y_ij}
\begin{aligned}
y^{h}_{\mathrm{p}, ij} (f) =&  \frac{ \hat{n}_{ij} \cdot {\mathrm{ e_p}} \cdot \hat{n}_{ij} }{2 (1 - \hat{n}_{ij} \cdot \hat{k} ) } 
 \times \left[  \exp( 2 \pi i f (L_{ij} + \hat{k} \cdot p_i ) ) -  \exp( 2 \pi i f  \hat{k} \cdot p_j )  \right] ,
\end{aligned}
\end{equation}
\end{widetext}
where $\hat{n}_{ij}$ is the unit vector from S/C$i$ to $j$, $L_{ij}$ is the arm length from S/C$i$ to $j$, $p_i$ is the position of the S/C$i$ in the solar-system barycentric ecliptic coordinates.

The first-generation Michelson TDI configuration and its corresponding optimal channels are employed to represent the performance of each mission. The response of the Michelson-X channel for a specific polarization p in the frequency domain will be the sum of the responses in the time shift single links,
\begin{equation} \label{eq:resp_X_FD}
\begin{aligned}
 F_{ \rm X,p} (f) =& (-\Delta_{21} + \Delta_{21}  \Delta_{13}  \Delta_{31})  y^{h}_\mathrm{p,12} \\
         & + (-1 + \Delta_{13}  \Delta_{31} )  y^{h}_\mathrm{p,21}  \\
         & + (\Delta_{31} - \Delta_{31}  \Delta_{12}  \Delta_{21})  y^{h}_\mathrm{p,13} \\
         & + ( 1 - \Delta_{12}  \Delta_{21} )  y^{h}_\mathrm{p,31}, \\
\end{aligned}
\end{equation}
where $\Delta_{ij} = \exp(2 \pi i f L_{ij})$. The GW responses in the Michelson optimal A, E, and T channels are obtained by applying \cite{Prince:2002hp,Vallisneri:2007xa}
\begin{equation} \label{eq:optimalTDI}
 {\rm A} =  \frac{ {\rm Z} - {\rm X} }{\sqrt{2}} , \quad {\rm E} = \frac{ {\rm X} - 2 {\rm Y} + {\rm Z} }{\sqrt{6}} , \quad {\rm T} = \frac{ {\rm X} + {\rm Y} + {\rm Z} }{\sqrt{3}},
\end{equation}
where Y and Z channels are obtained from cyclical permutation of the S/C indexes in the X channel. 

\subsection{The sensitivity of the networks}

\subsubsection{The noises in TDI}

Multiple noise sources will be involved in the process of TDI combinations from the link measurements. For the Michelson-X channels, the expression of measurements could be described as \cite{Vallisneri:2012np},
\begin{equation} \label{eq:X_measurement}
\begin{aligned}
{\rm X} =& [ \mathcal{D}_{31} \mathcal{D}_{13} \mathcal{D}_{21} \eta_{12}  + \mathcal{D}_{31}  \mathcal{D}_{13} \eta_{21}  +  \mathcal{D}_{31} \eta_{13} +  \eta_{31}   ] \\
& - [ \eta_{21} + \mathcal{D}_{21} \eta_{12} +\mathcal{D}_{21} \mathcal{D}_{12} \eta_{31} + \mathcal{D}_{21}  \mathcal{D}_{12} \mathcal{D}_{31} \eta_{13} ], \\
\end{aligned}
\end{equation}
where $\mathcal{D}_{ij}$ is a time-delay operator, $ \mathcal{D}_{ij} \eta(t) = \eta(t - L_{ij} )$, $\eta_{ji}$ are the combined observables from S/C$j$ to S/C$i$ \cite{Otto:2012dk,Otto:2015,Tinto:2018kij}, and the specific expressions for this work are defined in \cite{Wang:2ndTDI}.
By assuming the dominant laser frequency noises are sufficiently suppressed, the acceleration noise and optical path noise in $\eta_{ij}$ become the primary noise sources after TDI process. 

The noise budgets for the acceleration noise $S_{\rm acc}$ are assumed to be the same for the LISA and TAIJI \cite{2017arXiv170200786A,Luo:2020},
\begin{equation}
 S^{1/2}_{\rm acc} = 3 \times 10^{-15} \frac{\rm m/s^2}{\sqrt{\rm Hz}} \sqrt{1 + \left(\frac{0.4 {\rm mHz}}{f} \right)^2 }  \sqrt{1 + \left(\frac{f}{8 {\rm mHz}} \right)^4 }.
\end{equation}
And the optical path noises $S_{\rm op}$ requirement for two missions are treated slightly different as
\begin{equation}
\begin{aligned}
 S^{1/2}_{\rm op, LISA} & = 10 \times 10^{-12} \frac{\rm m}{\sqrt{\rm Hz}} \sqrt{1 + \left(\frac{2 {\rm mHz}}{f} \right)^4 },  \\
S^{1/2}_{\rm op, TAIJI} & = 8 \times 10^{-12} \frac{\rm m}{\sqrt{\rm Hz}} \sqrt{1 + \left(\frac{2 {\rm mHz}}{f} \right)^4 }.
 \end{aligned}
\end{equation}
And the power spectrum density (PSD) of a TDI channel $\mathrm{S}_\mathrm{n,TDI}$ is obtained by implementing the numerical algorithm developed in \cite{Wang:1stTDI,Wang:2ndTDI}.

\subsubsection{The joint sensitivities}

The antenna pattern of an interferometer will change with the geometric angles $\Omega(\lambda, \theta, \psi, \iota)$ and the frequency. For a given $\Omega$ and frequency, the sensitivities of the LISA's A+E+T channel and joint LISA-TAIJI network at a given mission time could be evaluated respectively by, 
\begin{align}
\mathrm{S}^{1/2}_{\rm LISA} (f, \Omega) &= \left( \sum_{\rm A,E,T} \frac{|F_{\rm TDI} (f, \Omega) |^2}{ \mathrm{S}_{\rm n, TDI} (f) }  \right)^{-1/2}, \label{eq:S_LISA} \\
\mathrm{S}^{1/2}_{\rm joint} (f, \Omega) &= \left( \sum^{\rm TAIJI}_{\rm LISA} \sum_{\rm A,E,T} \frac{|F_{\rm TDI} (f, \Omega )|^2}{ \mathrm{S}_{\rm n, TDI} (f) } \right)^{-1/2} \label{eq:S_LISA_TAIJI}.
\end{align}

\begin{figure*}[htb]
\includegraphics[width=0.48\textwidth]{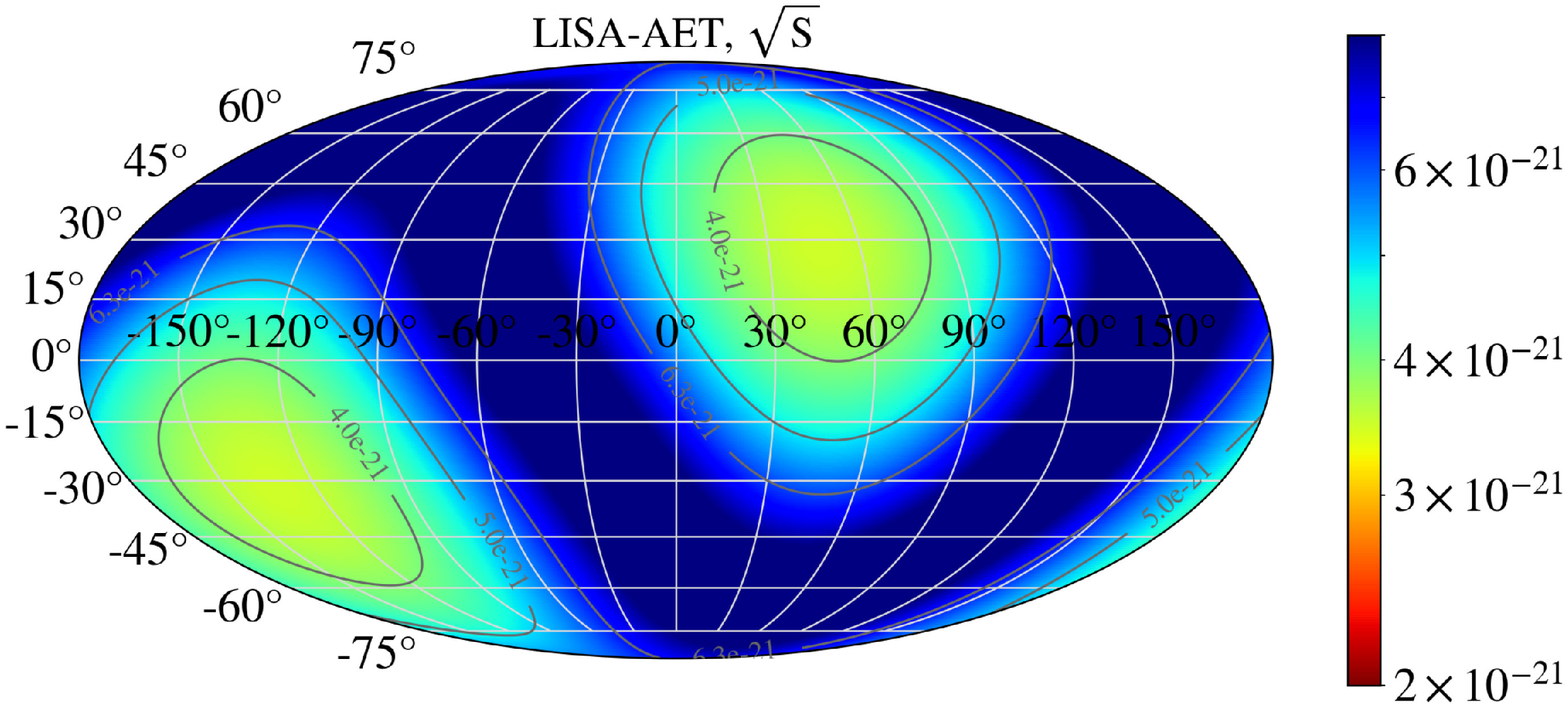} 
\includegraphics[width=0.48\textwidth]{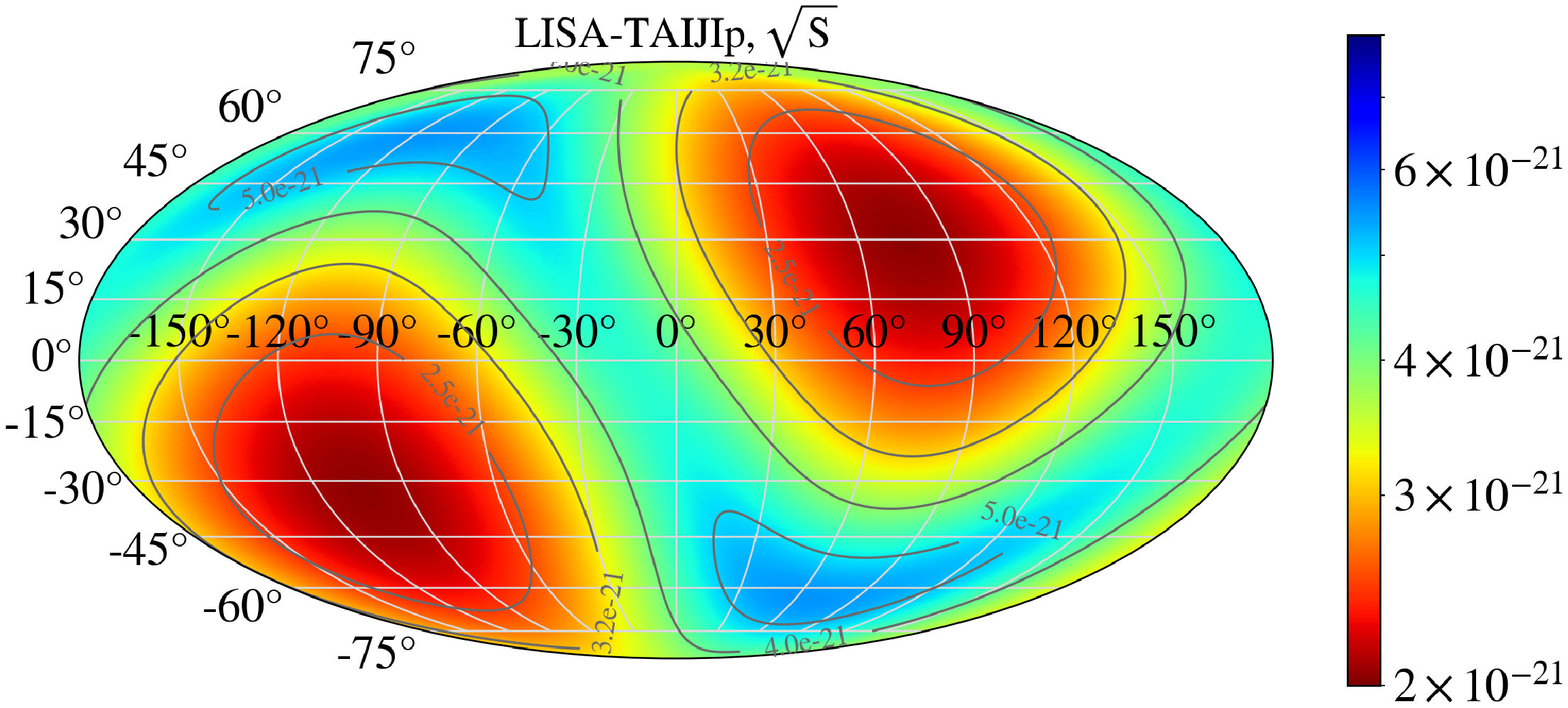} 
\includegraphics[width=0.48\textwidth]{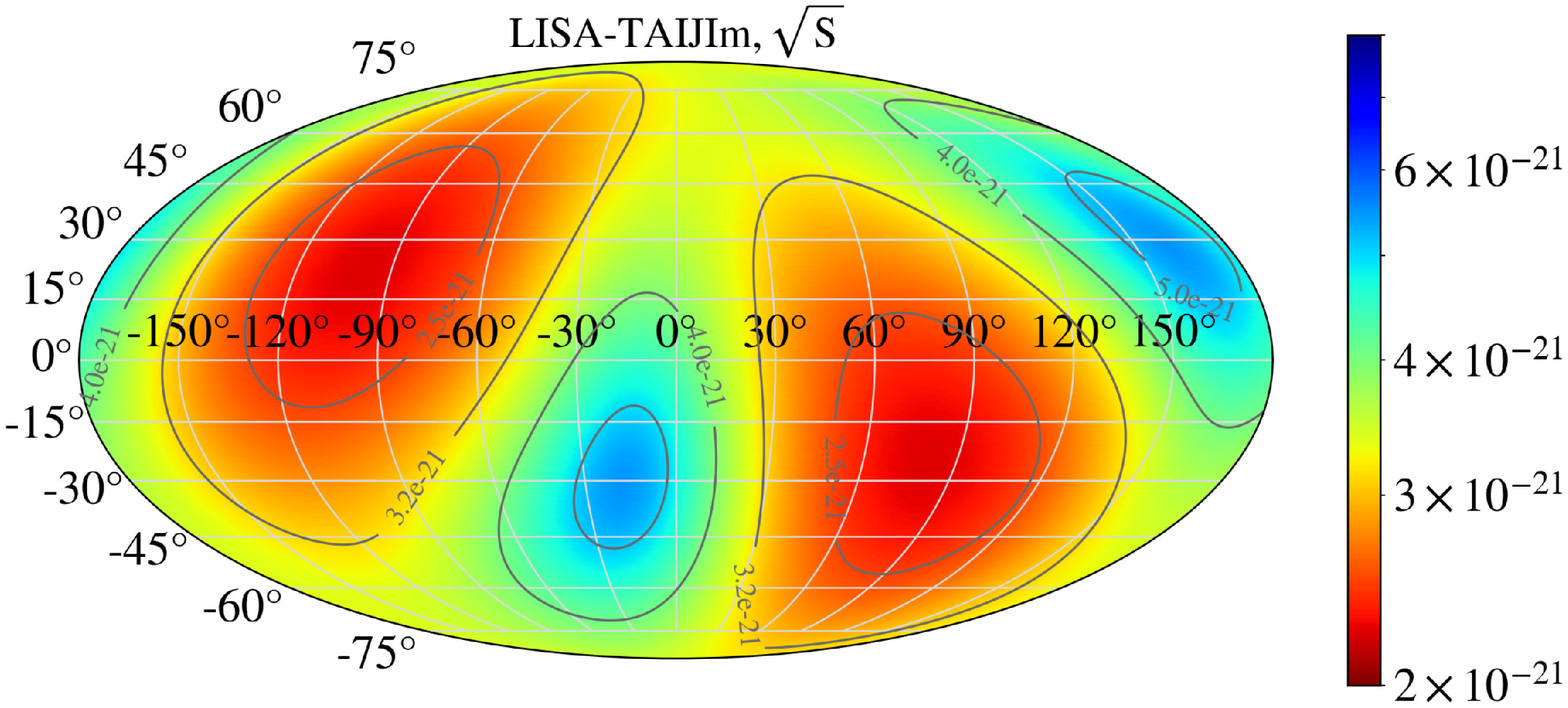} 
\includegraphics[width=0.48\textwidth]{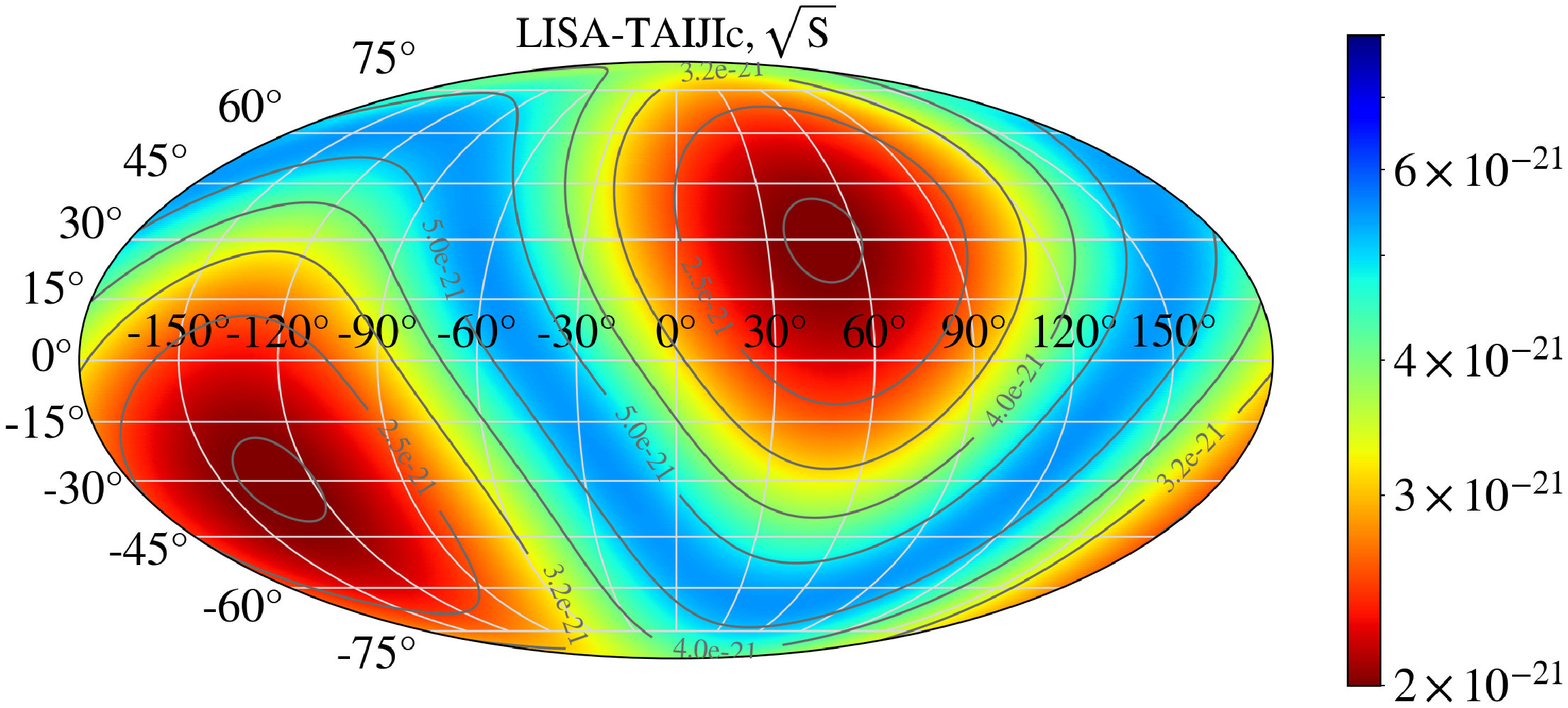} 
\caption{\label{fig:LISA_TAIJI_sensitivity_Mollweide} The instantaneous sensitivities on the sky map for the LISA mission and LISA-TAIJI networks at $\psi=0, \iota=0$, and $f = 10$ mHz. The single LISA sensitivity is shown by the upper left panel which obtained by using Eq. \eqref{eq:S_LISA}, and the joint LISA-TAIJI network sensitivities are obtained by using Eq. \eqref{eq:S_LISA_TAIJI}. The sensitivity of the LISA-TAIJIp is shown in the upper right panel, the sensitivity of LISA-TAIJIm is shown in the lower left panel, and the sensitivity of the LISA-TAIJIc is shown by the lower right panel. The plots reflect the antenna pattern of the detectors considering the orientation of the S/C formations. The LISA-TAIJIm network achieves a better averaged sensitivity to the different sky directions than the other two networks. (One caveat is that the sensitivity of the TAIJI is slightly better than the LISA mission, fully symmetry on the plots should not be expected).}
\end{figure*}

The instantaneous sensitivities to the tensor polarizations for the LISA and joint LISA-TAIJI networks for $\psi=0, \iota=0,$ and $f = 10 \  \mathrm{mHz}$ are shown in Fig. \ref{fig:LISA_TAIJI_sensitivity_Mollweide}. As we can see in the upper left plot, the LISA has the optimum sensitivity around the normal directions (ecliptic latitude $\pm 30^\circ$ ) of the triangular formation plane considering its $60^\circ$ inclination. As expected from Fig. \ref{fig:LISA_TAIJI}, the antenna pattern of the TAIJIp is shifted by $\sim 40^\circ$ along the ecliptic latitude with respect to the LISA's. And their joint sensitivity is shown by the upper right panel in Fig. \ref{fig:LISA_TAIJI_sensitivity_Mollweide}.
For the TAIJIm, due to its $40^\circ$ separation and $-60^\circ$ inclination with respect to the LISA, its antenna pattern is not only shifted by $40^\circ$ along the latitude, also inversed with respect to the ecliptic plane. And their joint sensitivity of the LISA-TAIJIm is shown in the lower left panel. As for the TAIJIc case, since the TAIJIc is co-located and coplanar with LISA, the joint LISA-TAIJIc enhanced the LISA's sensitivity as shown in the lower right plot. One caveat is that the sensitivity of the TAIJI is slightly better than the LISA mission, fully symmetry should not be expected for the joint sensitivity plots in Fig. \ref{fig:LISA_TAIJI_sensitivity_Mollweide}.

\section{Parameter determinations for SMBH binary Coalescence} \label{sec:SMBH_results}

As the most promising GW source for the LISA and TAIJI missions, the SMBH binary is selected to demonstrate the performances of parameter determination from the LISA and three LISA-TAIJI networks. 

\subsection{Fisher information method}

The Fisher information matrix (FIM) is employed in this investigation to determine the uncertainty of parameters from GW observation \cite[and references therein]{1994PhRvD..49.2658C,Cutler:1997ta,Vallisneri:2007ev,Kuns:2019upi}. 
For a single mission with full six links, the FIM is combined from three optimal channels (A, E, and T). And the FIM of the joint LISA-TAIJI network is obtained by summing up the FIM from two missions,
\begin{equation}
\Gamma_{ij}  = \sum^{\mathrm{TAIJI}}_{\rm LISA} \sum_{\rm A,E,T}  \left( \frac{\partial \tilde{h}_\mathrm{TDI} }{ \partial {\xi_i} } \bigg\rvert \frac{\partial \tilde{h}_\mathrm{TDI} }{ \partial {\xi_j} }  \right),
\end{equation}
with
\begin{equation}
\left( g | h\right)_{\rm TDI} = 4 \mathrm{Re} \int^{\infty}_0 \frac{g^{\ast} (f) h(f)}{S_{\rm TDI} (f) } \mathrm{d} f ,
\end{equation}
where $\tilde{h}_\mathrm{TDI}$ is the frequency domain GW waveform responded in a TDI channel, $\xi_i$ is the $i$-th parameter to be determined, and $S_{\rm TDI} (f)$ is the noise PSD of the corresponding TDI channel. 

Considering the source location estimation will be significantly affected by the polarization content of the source \cite{Abbott:2017oio,GW170817:detection,GW170817:TGR}, only tensor polarizations from GR are included to investigate the angular resolution of the sky localization. And 9 parameters are utilized to describe the GW signal and TDI responses from the LISA or LISA-TAIJI network, which are ecliptic longitude and latitude $(\lambda, \theta)$, polarization angle $\psi$, source inclination $\iota$, luminosity distance $D$, the coalescence time and phase $(t_c, \phi_c)$, the total mass of binary $M$ and mass ratio $q$. The GW signal responded by TDI incorporating two polarizations ($+$ and $\times$) could be described as
\begin{equation} \label{eq:h_FD_GW}
\begin{aligned}
\tilde{h}_\mathrm{GR,TDI} (f) = & ( F_{+} + F_{\times} ) \tilde{h}_\mathrm{GR}  ,
\end{aligned}
\end{equation}
where $\tilde{h}_\mathrm{GR}$ is the frequency domain waveform represented by IMRPhenomPv2 \cite{Khan:2015jqa}.
When the constraints on the alternative GW polarizations are investigated, additional 6 ppE (parametrized post-Einsteinian) parameters, ($\beta, b, \alpha_\mathrm{b}, \alpha_\mathrm{L}, \alpha_\mathrm{x}, \alpha_\mathrm{x} $), are employed to qualify the deviations from the GR as developed in \cite{yunes2012PhRvD..86b2004C}, and the waveform will be explained in Eq. \eqref{eq:h_FD_ppE}.

The variance-covariance matrix of the parameters is calculated by
\begin{equation}
\begin{aligned}
 \left\langle \Delta \xi_i \Delta \xi_j  \right\rangle = \left( \Gamma^{-1} \right)_{ij} + \mathcal{O}({\rho}^{-1})
\overset{{\rho} \gg 1}{\simeq } \left( \Gamma^{-1} \right)_{ij} . \\
\end{aligned}
\end{equation}
The standard deviations $\sigma_i$ and correlation $\sigma_{ij}$ of the parameters for the high SNR $\rho \gg 1$ will be
\begin{equation}
\begin{aligned}
\sigma_{i} & \simeq \sqrt{ \left( \Gamma^{-1} \right)_{ii}  }, \\
\sigma_{ij} & = \mathrm{cov} (\xi_i, \xi_j)  \simeq \left( \Gamma^{-1} \right)_{ij} .
\end{aligned}
\end{equation}
The uncertainty of sky localization for one source is evaluated by 
\begin{equation} \label{eq:delta_Omega}
 \Delta \Omega \simeq 2 \pi | \cos \theta | \sqrt{ \sigma_{\lambda } \sigma_{\theta } -  \sigma^2_{\lambda \theta}   }.
\end{equation}

The Monte Carlo simulation is performed for parameter determination by 1000 sources. The $(\lambda, \theta)$ are randomly sampled in the sky sphere, $\psi$ is sampled in $[0, 2 \pi]$ uniformly, $\cos \iota$ is sampled randomly in $[-1, 1]$, the merge time $t_c$ is randomly in one year. The $m_1 = 10^5\ M_{\odot}, q=1/3$ at redshift $z=2$ is fixed as we used in \cite{Wang:2020a,Wang:2021polar}. Considering the SNR is mainly contributed from the binary coalescing stage, the 30 days observation before the merge is simulated to perform the investigation.

\subsection{Sky localization of the networks} 

\begin{figure*}[htb]
\includegraphics[width=0.46\textwidth]{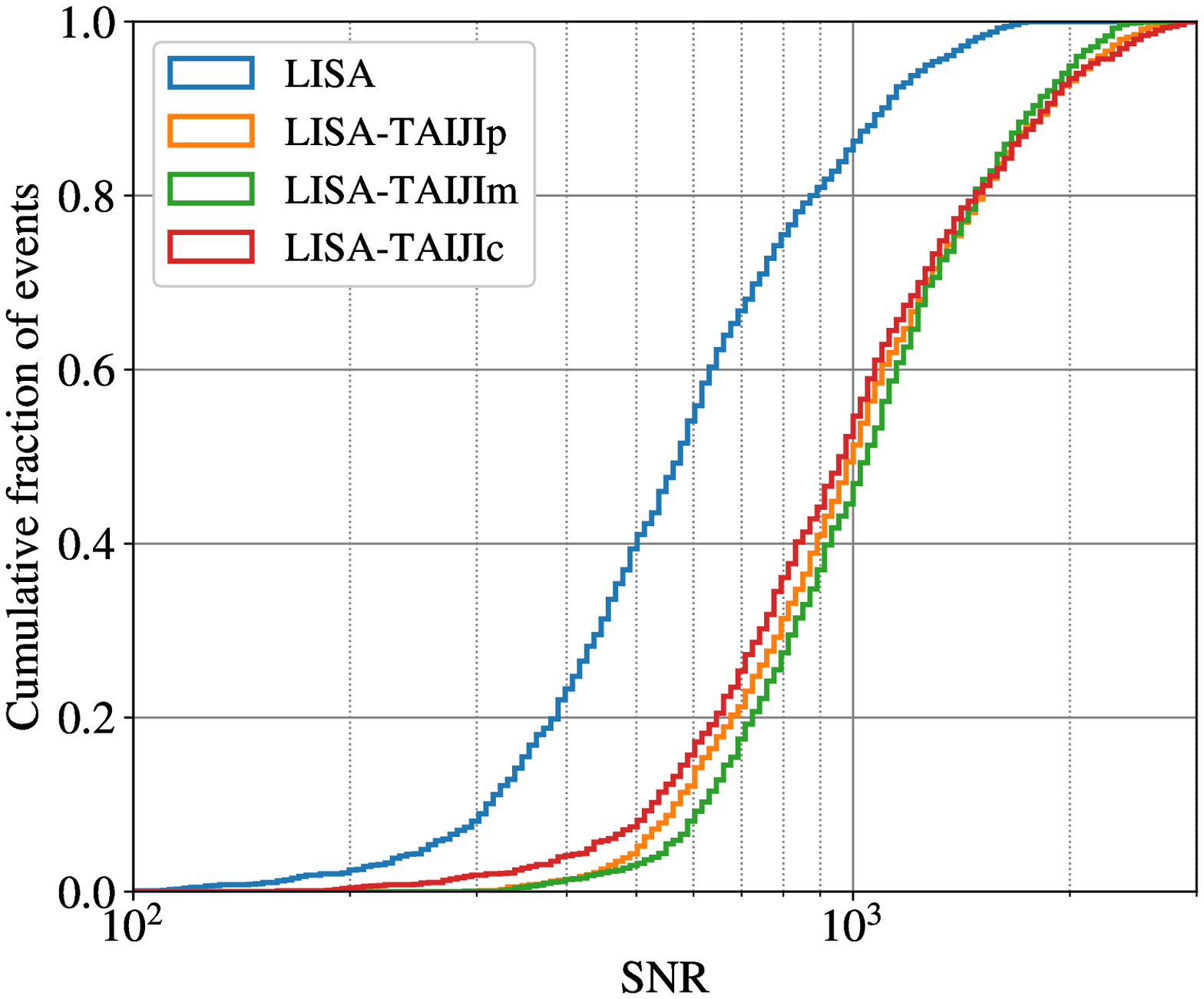} 
\includegraphics[width=0.46\textwidth]{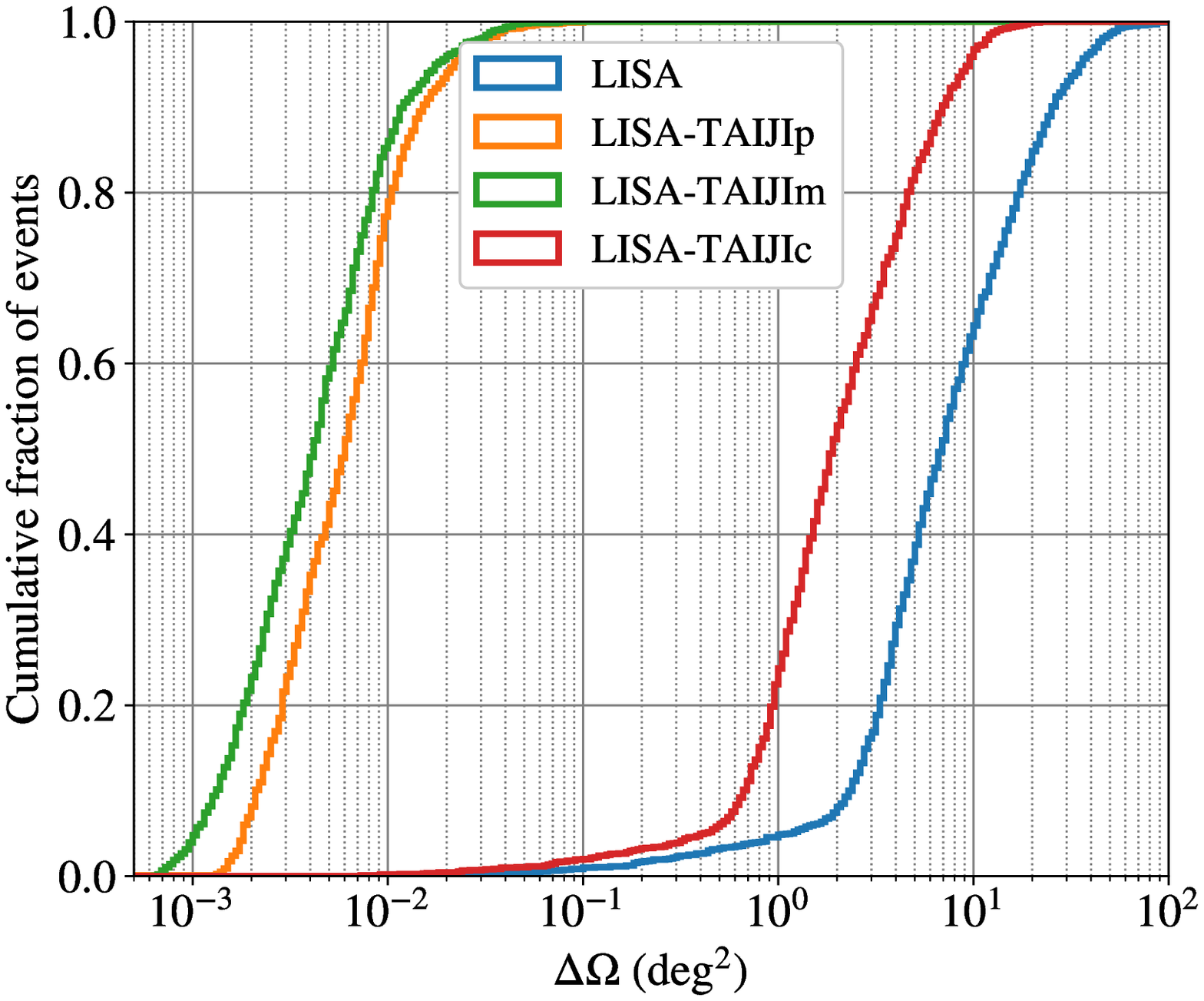} 
\caption{The cumulative histograms of SNR (left panel) and angular resolutions of the sky localization (right panel) from the LISA and three LISA-TAIJI networks. Three LISA-TAIJI networks could achieve more than $ \sqrt{2}$ times SNR of the LISA mission. In the three networks, the SNR from LISA-TAIJIm has the most concentrated distribution because of the most averaged antenna pattern as shown in Fig. \ref{fig:LISA_TAIJI_sensitivity_Mollweide}. The distribution of SNR from LISA-TAIJIc has a long tail since the two missions share the same sensitive/insensitive areas. On the right plot, the resolution of sky localization by LISA-TAIJIc is approximately twice better than the single LISA as the attribute to the SNR increase. The angular resolution of the LISA-TAIJIp is better than LISA-TAIJIc benefiting from the long separation. The performance of LISA-TAIJIm is better than LISA-TAIJIp because the angle of the LISA and TAIJIm formation planes is $71^\circ$ and their orientation could yield more collaborative antenna pattern. \label{fig:hist_SNR_sky_localization} 
}
\end{figure*}

The cumulative histograms of SNR from the LISA and LISA-TAIJI networks are shown in the left panel of Fig. \ref{fig:hist_SNR_sky_localization}. Compared to the single LISA mission, three LISA-TAIJI networks achieve more than $\sqrt{2}$ SNR by implementing the quadratic sum ($\rho^2_\mathrm{joint}=\rho^2_\mathrm{LISA} + \rho^2_\mathrm{TAIJI}$) considering the TAIJI is more sensitive to LISA in the selected GW frequency band. In the three networks, the LISA-TAIJIc has a larger range of the SNR distribution with a longer tail, because the co-aligned detectors are sensitive/insensitive to the same directions and leaving the common optimal/blind areas. The LISA-TAIJIm shows the most concentrated SNRs values compared to the two other networks since their joint antenna pattern is more averaged on the sky map as shown in Fig. \ref{fig:LISA_TAIJI_sensitivity_Mollweide}.
 
The angular resolutions of the sky localization from the LISA and LISA-TAIJI networks are shown in the right panel of Fig. \ref{fig:hist_SNR_sky_localization}. 
For the LISA-TAIJIc network, the uncertainties of sky localization are improved by more than 2 times compared to the single LISA mission which should be due to the more than $\sqrt{2}$ times SNR from the network. Compared to the LISA-TAIJIc, the joint observation from LISA-TAIJIp demonstrate the more than 2 orders improvement on the localization resolution which should be mainly attributed to the long baseline separations between the LISA and TAIJIp. 
On the other side, the LISA-TAIJIm yield a better capability on locating the source than LISA-TAIJIp because the TAIJIm's formation plane is $71^\circ$ with respect to LISA's, and its antenna pattern could better compensate the LISA's insensitive directions. 

\subsection{Observation for GW polarizations}
 
The detector responded GW signal is modified as follows to incorporate alternative polarization beyond GR \cite{yunes2012PhRvD..86b2004C,Wang:2021polar},
\begin{equation} \label{eq:h_FD_ppE}
\begin{aligned}
\tilde{h}_\mathrm{ppE, TDI} (f) = &  \left[ ( F_\mathrm{TDI,+} + F_\mathrm{TDI,\times} ) ( 1 + C \beta u^{b+5}_2 ) \right.  \\
& \left. + \alpha_\mathrm{b} F_\mathrm{TDI,b} +  \alpha_\mathrm{L} F_\mathrm{TDI,L} \right. \\
& \left. + \alpha_\mathrm{x} F_\mathrm{TDI,x} + \alpha_\mathrm{y} F_\mathrm{TDI,y}   \right] \tilde{h}_\mathrm{GR} \ e^{ 2 i \beta u^b_2} ,
\end{aligned}
\end{equation} 
where $C$ is the function of $b$ and is defined by Eq. (11) in the Erratum \cite{PhysRevD.95.129901} of \cite{yunes2012PhRvD..86b2004C}, $\tilde{h}_\mathrm{GR}$ is the GW waveform represented by IMRPhenomPv2 \cite{Khan:2015jqa}, and $u_2 \equiv (\pi \mathcal{M}f)^{1/3}$. In this investigation, we choose $b=-3$ which corresponds to the massive graviton theory \cite{Will:1997bb,Will:2004xi,Berti:2005qd,Stavridis:2009mb,Arun:2009pq,Keppel:2010qu,Yagi:2009zm}, and set $\beta=0.01$ which is from the rough boundary constrained in \cite{Cornish:2011ys}.
The other four ppE parameters tuning the amplitudes of the alternative polarizations are set to be zeros, $(\alpha_\mathrm{b}$, $\alpha_\mathrm{L}$, $\alpha_\mathrm{x}$, $\alpha_\mathrm{y}) = (0,0,0,0)$. 
Although this specific selection could not represent all other gravity theories, we have demonstrated that the measurement on the ppE parameters could also be similarly improved by the LISA-TAIJI network for other value choices \cite{Wang:2021polar}.

\begin{figure*}[htb]
\includegraphics[width=0.46\textwidth]{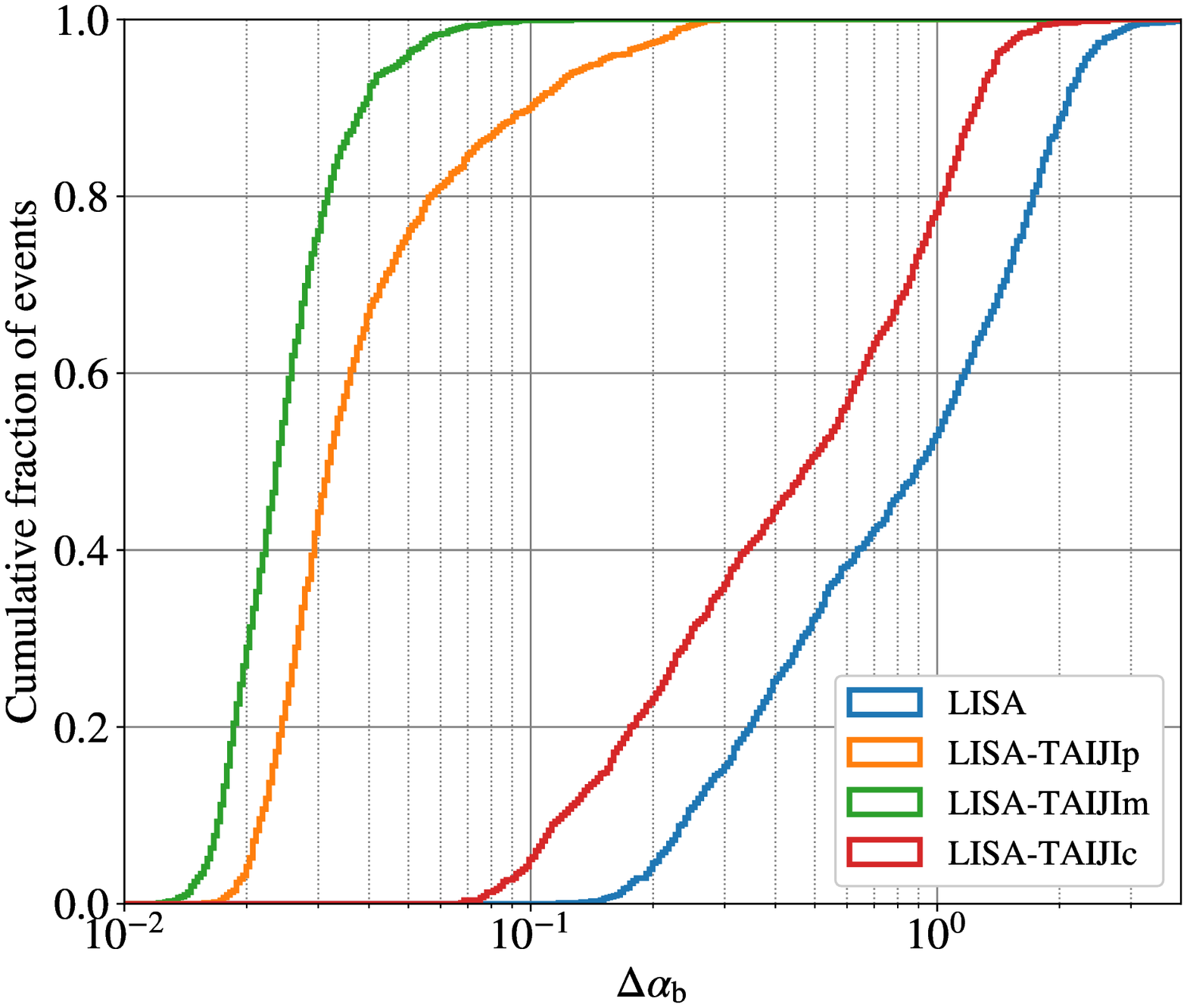} 
\includegraphics[width=0.46\textwidth]{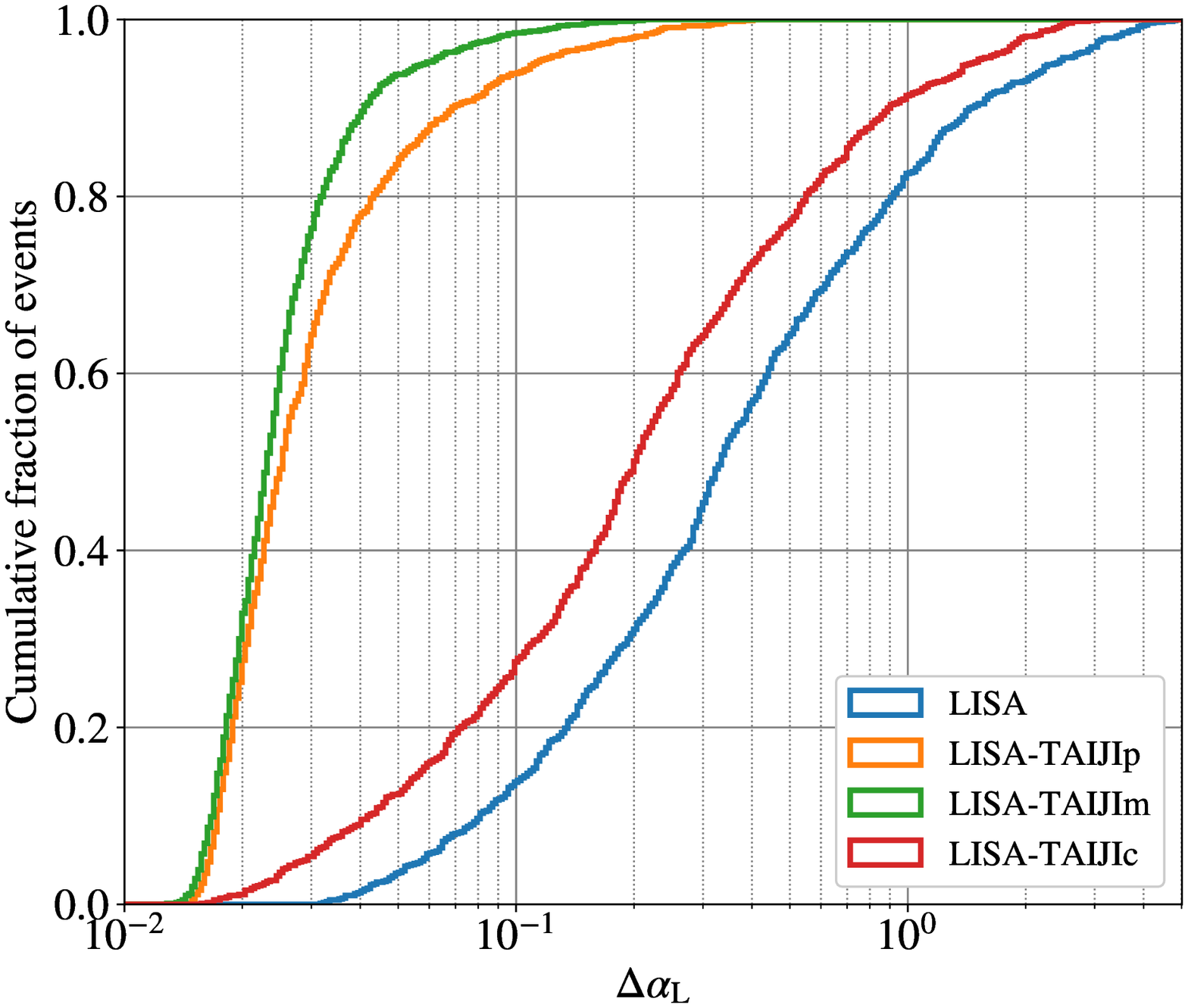} 
\includegraphics[width=0.46\textwidth]{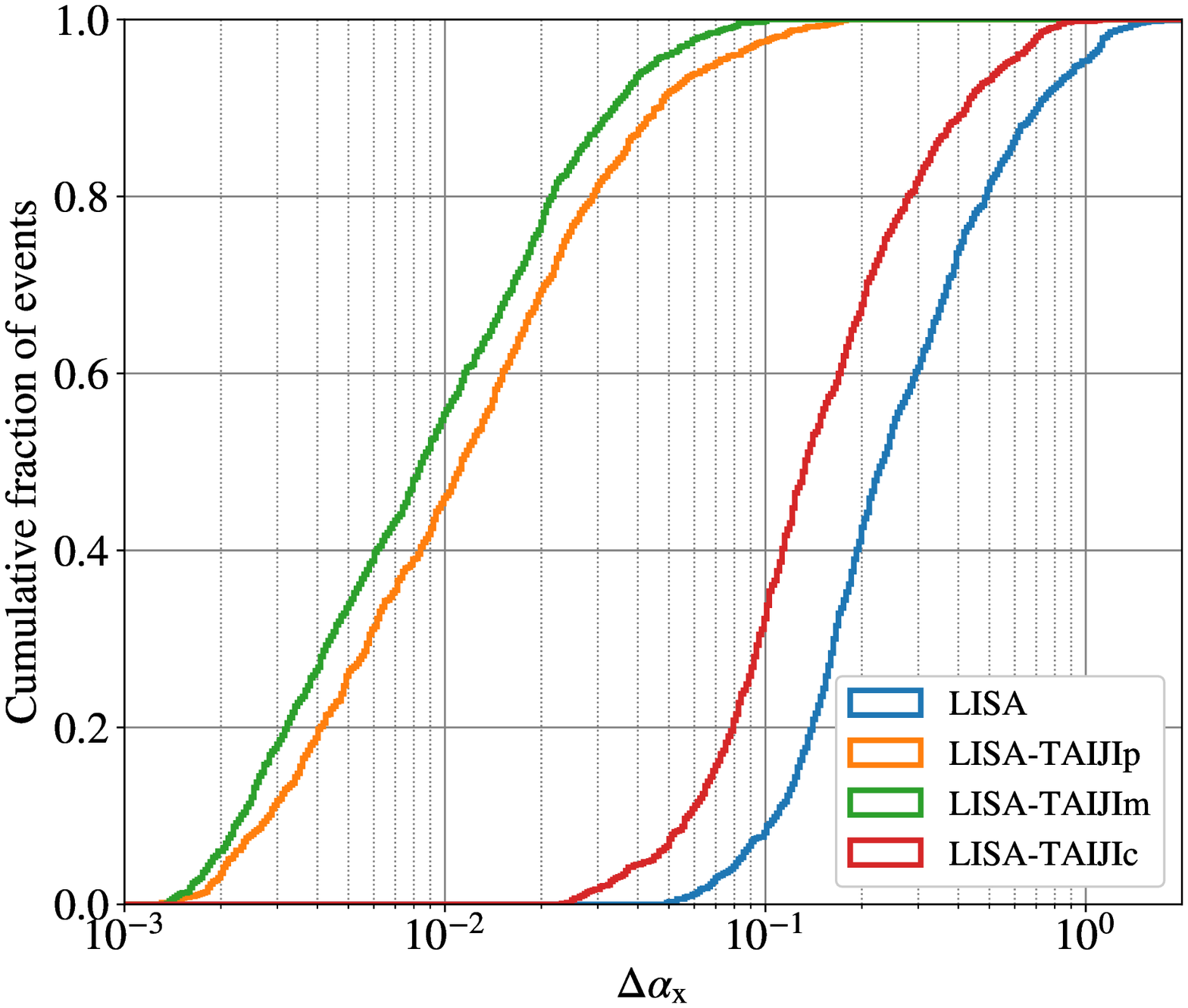} 
\includegraphics[width=0.46\textwidth]{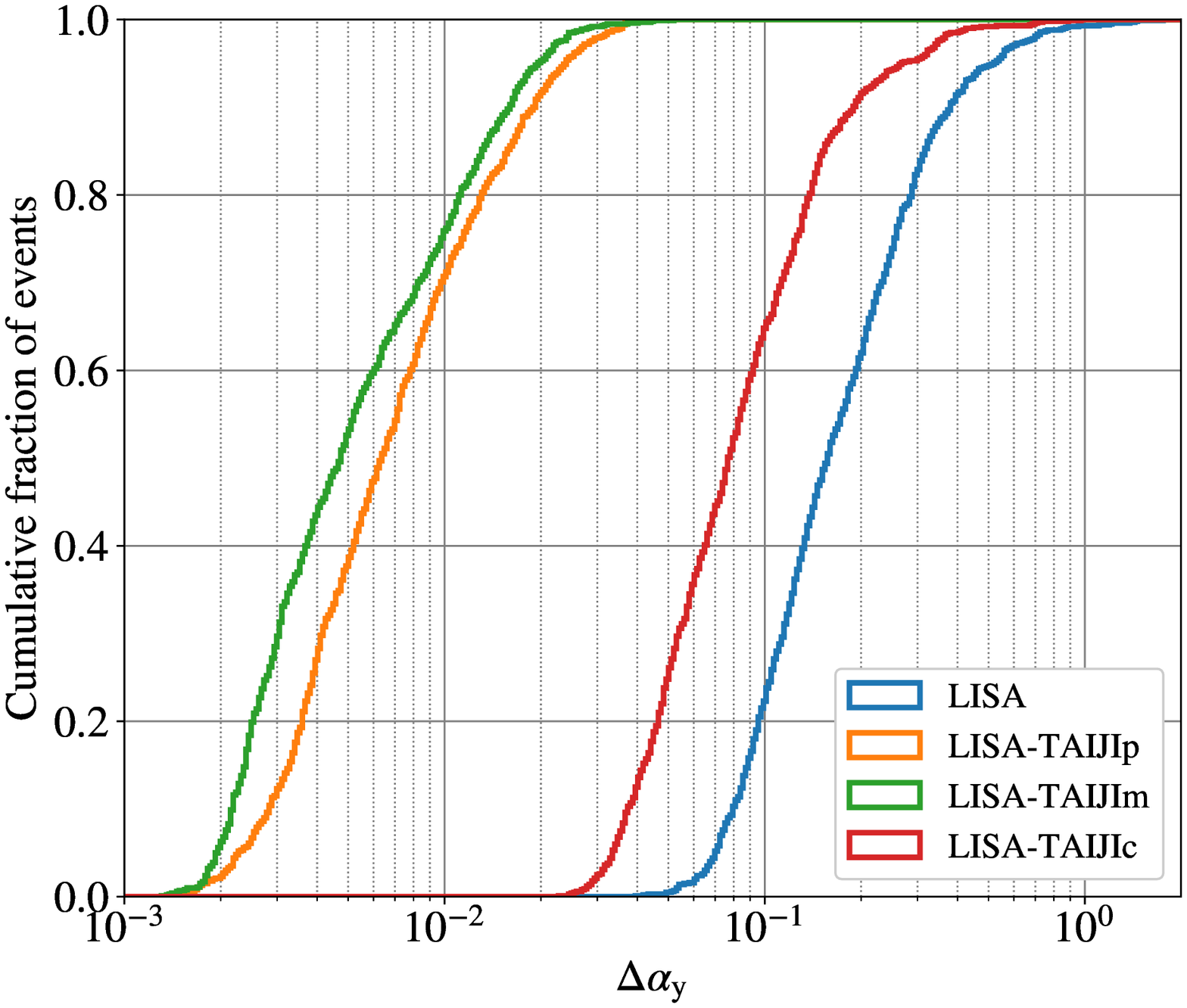} 
\caption{The cumulative histograms of the constraints on the amplitudes of scalar and vector polarizations. The result from the LISA-TAIJIc is more than $\sqrt{2}$ times better than the single LISA which is an attribute from the increase of SNR. The LISA-TAIJIp, with long-baseline separation, could significantly improve the constraints on polarizations. And LISA-TAIJIm achieves the best constraints on the polarizations in three LISA-TAIJI configurations as the benefit of better antenna pattern cooperation. \label{fig:hist_polarizations}
}
\end{figure*}

The constraints on the ppE parameters $\alpha$ for the scalar (upper panel) and vector (lower panel) polarizations are shown in Fig. \ref{fig:hist_polarizations}. Similar to the results achieved for the sky localization, the constraint on $\alpha$ from the LISA-TAIJIc is more than $\sqrt{2}$ times better than single LISA which is an attribute from the increase of SNR. The LISA-TAIJIp, with large separation, could improve the constraints on the polarizations significantly. And LISA-TAIJIm could achieve the best constraints on the polarizations in the three LISA-TAIJI configurations as the benefit of better antenna pattern cooperation.

\section{Overlap reduction function of the LISA-TAIJI networks} \label{sec:stochastic_background}

The response of the detector network to the stochastic background GW signal will depend on the locations and orientations of the interferometers. \citet{Flanagan:1993ix} evaluated the sensitivities of the ground-based GW interferometers to the stochastic background. An \textit{overlap reduction function} is introduced to indicate the cross-correlation between a pair of detectors \cite{Christensen:1992wi}. \citet{Whelan:2001qw} calculated the overlap reduction functions for the two LIGO detectors and GEO. \citet{Omiya:2020fvw,Seto:2020mfd}, and \citet{Orlando:2020oko} specified the overlap function of the LISA-TAIJIp network for optimal TDI channels and alternative GW polarizations. \citet{Schmitz:2020syl} reviewed the detectability of the ground- and space-based detectors for the stochastic GW background. 

\begin{figure*}[htb]
\includegraphics[width=0.46\textwidth]{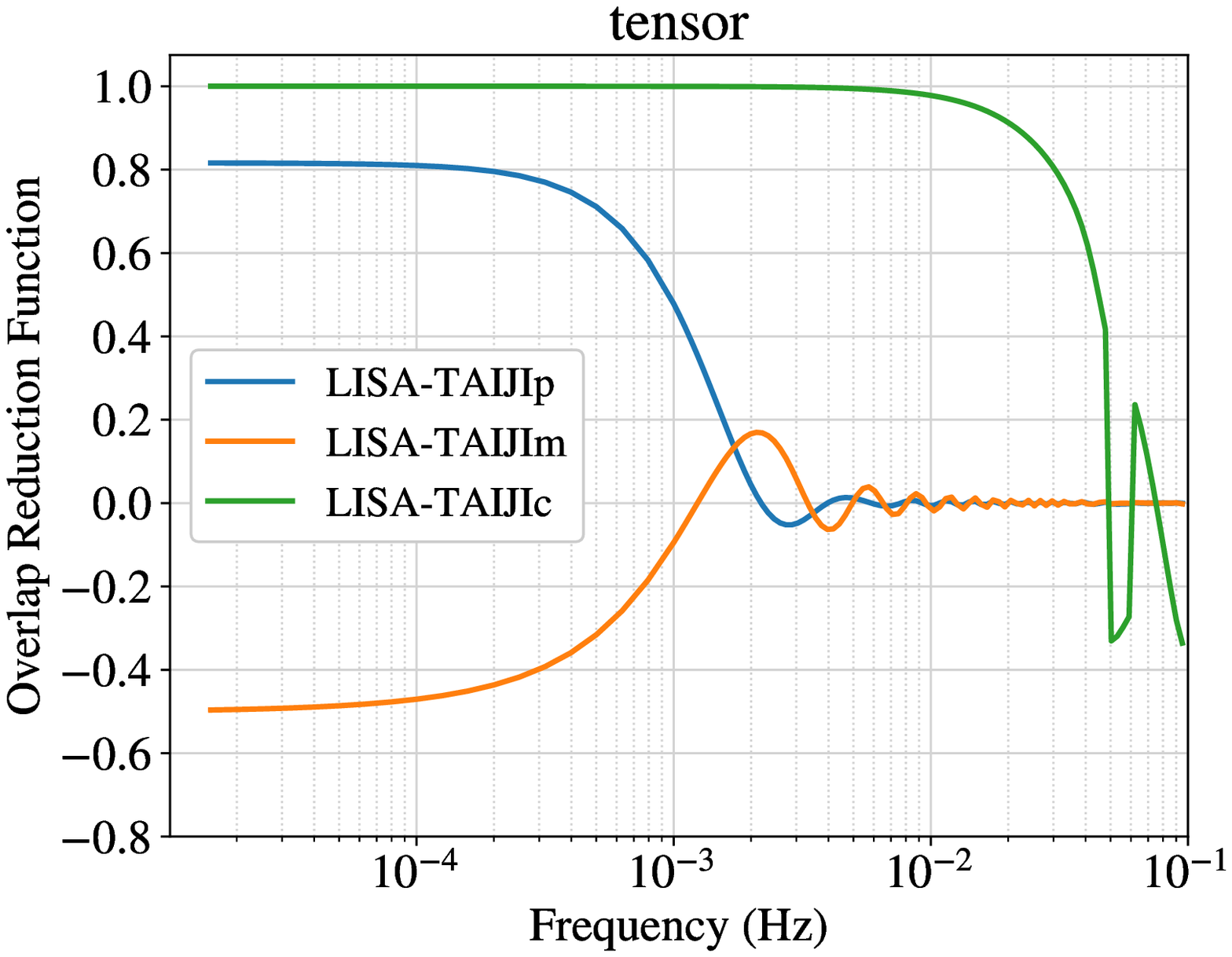} 
\includegraphics[width=0.46\textwidth]{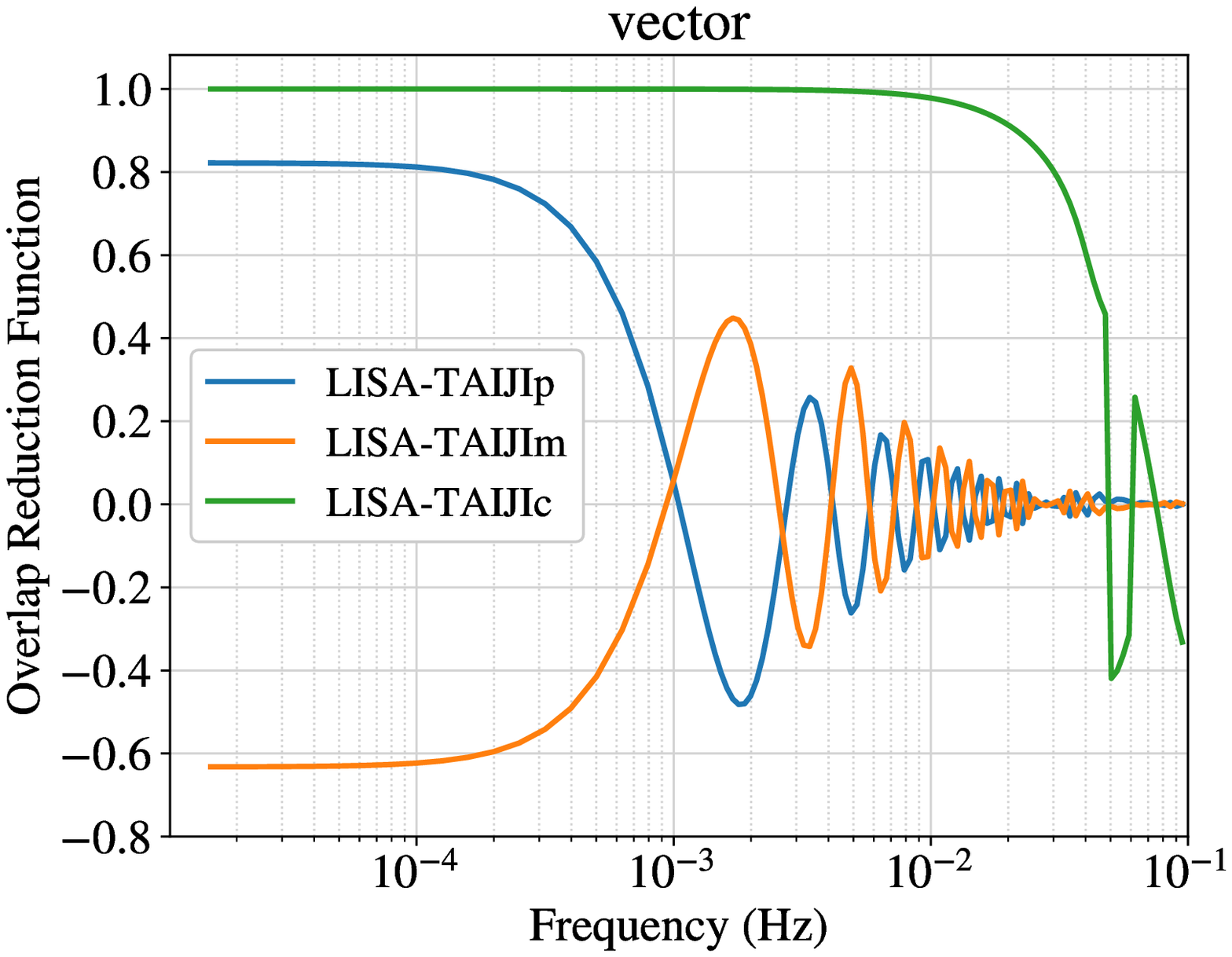} 
\includegraphics[width=0.46\textwidth]{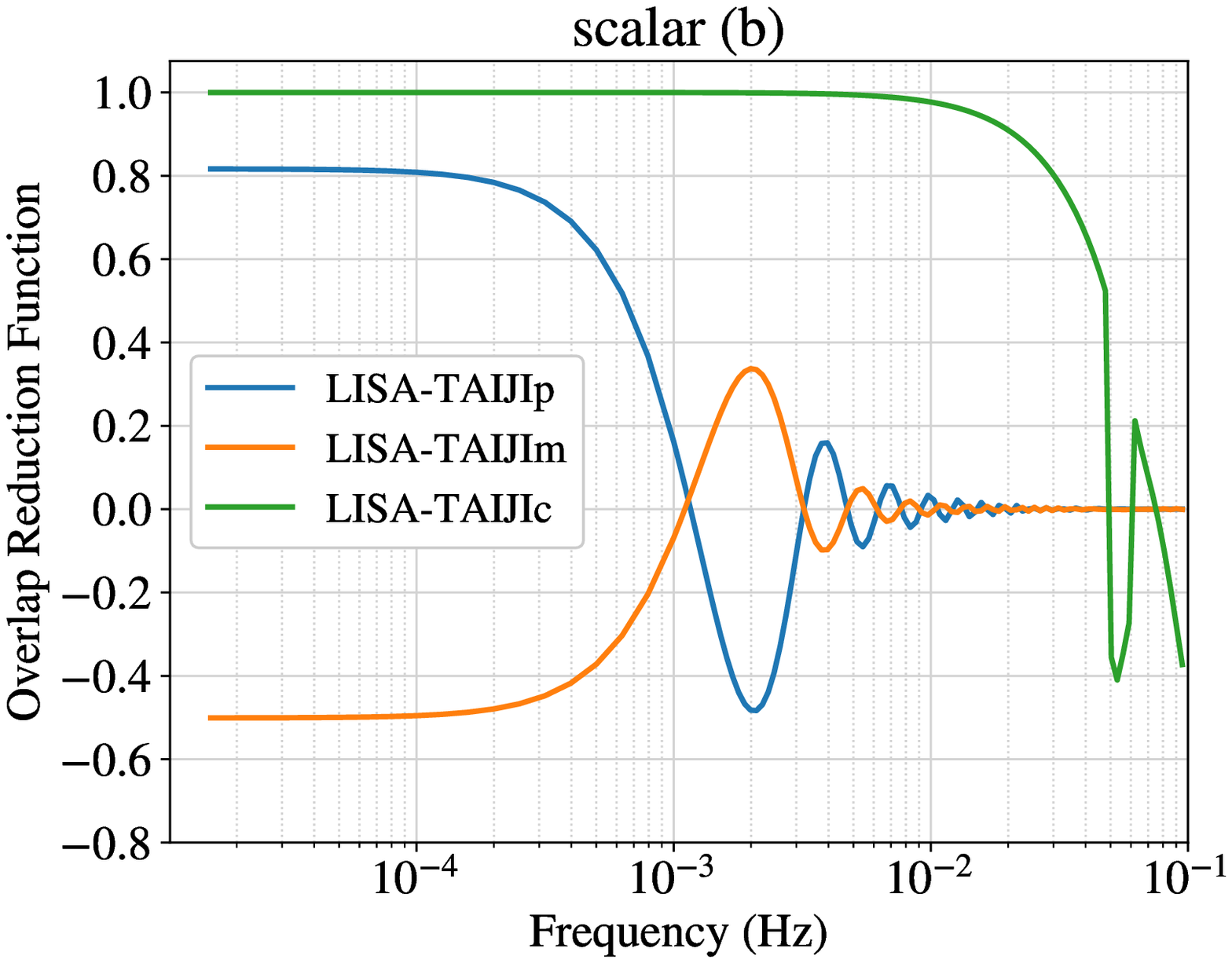} 
\includegraphics[width=0.46\textwidth]{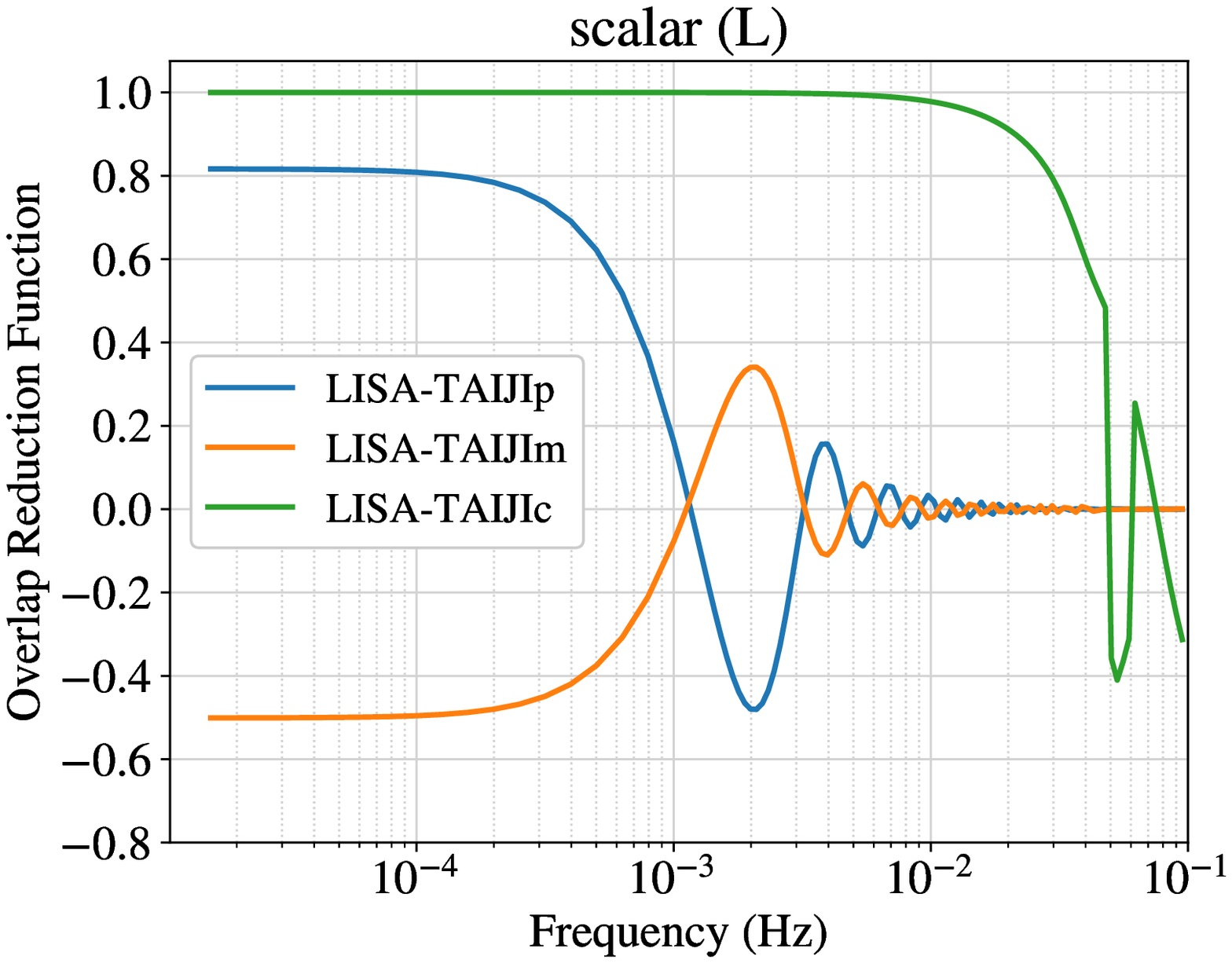} 
\caption{The overlap reduction functions of three LISA-TAIJI networks for different polarization modes. The overlap function of the LISA-TAIJIc network is unity for the frequency lower than 10 mHz which is optimal for the SGWB observation, and it changes the sign during the two detectors' characteristic frequencies gap [50 mHz, 60 mHz]. The overlap function of the LISA-TAIJIp/TAIJIm are close to zero around 1.5 mHz due to their $1 \times 10^8$ km separations ($f_\mathrm{crit} = c/(2 \times 1 \times 10^8 \ \mathrm{km}) \simeq 1.5 $ mHz). The LISA-TAIJIm network has a worse cross-correlation than the LISA-TAIJIp because of the more misaligned orientation.  \label{fig:overlap_reduction_fn} 
}
\end{figure*}

For a single LISA-like mission with full six measurement links, the optimal TDI channels could be treated as three equivalent interferometers. The observation from these TDI channels could be used to discriminate the stochastic GW background from the instrument noise \cite{Adams:2010vc}. And the motion of the detectors may also help to resolve the background, especially for the anisotropic signal \cite{Romano:2016dpx}. The LISA and TAIJI could form an ideal network to separate the cosmological SGWB signal from other stochastic processes such as the instrument noise and astrophysical foreground.   

To characterize the cross-correlation between LISA and different TAIJI orbital configurations, their overlap reduction functions are calculated for different polarizations,
\begin{equation}
\gamma_\mathrm{ab,p} (f) = \frac{\kappa}{4 \pi} \int  \mathrm{d} \mathbf{n}  \sum_\mathrm{A,E,T} F^{\mathrm{a}}_\mathrm{TDI,p} (f, \mathbf{n}) 
 \sum_\mathrm{A,E,T} F^\mathrm{b}_\mathrm{TDI,p} (f, \mathbf{n}) 
\end{equation}
where $F^{\mathrm{a}}_\mathrm{TDI,p}$ is the response function to the polarization mode p (tensor, vector, scalar breathing, and scalar longitudinal) in the TDI channel from the mission a, and $\kappa$ is the normalization factor to make $\gamma_\mathrm{ab}=1$ when the two detectors are co-aligned and co-localized. The polarization angle $\psi$ is set to be zero, and the inclination $\iota$ is set to be optimal for each polarization mode.

The overlap reduction functions for the different LISA-TAIJI networks for different polarizations are shown in Fig. \ref{fig:overlap_reduction_fn}.
Since the orientations of the LISA and TAIJIc are aligned and locations of them are coplanar and at the same location, the overlap function of the LISA-TAIJIc network is unity for the frequency lower than 10 mHz which indicates the strong correlation between the LISA and TAIJIc detectors, and the network is optimal for the SGWB observation for all polarization modes. We also notice that their overlap reduction functions change the sign during the two detectors' characteristic frequencies gap [$\frac{c}{2 L_\mathrm{TAIJI}} = 50$ mHz, $\frac{c}{2 L_\mathrm{LISA}} = 60$ mHz] ($c$ is the speed of the light in this section). 

Considering the $1 \times 10^8$ km separation between LISA and TAIJIp/TAIJIm, we can see the overlap reduction functions quickly approach to zero around a critical frequency $f_\mathrm{crit} \simeq c/(2 \times 1 \times 10^8 \ \mathrm{km}) \simeq 1.5 $ mHz \cite{Romano:2016dpx}, and oscillate and decay with the frequency increase. The $\gamma_\mathrm{ab}$ from the LISA-TAIJIp pair is higher than the value of the LISA-TAIJIm because the orientation of the LISA is more aligned with the TAIIJIp $(34.5^\circ)$ than the TAIJIm $(71^\circ)$ to make TAIJIp have a stronger correlation with the LISA. Therefore, in the three LISA-TAIJI networks, the LISA-TAIJIm should be the relatively worst configuration for the SGWB detections.

There will be a trade-off for a LISA-like GW detector network to observe the compact binary system and SGWB. A long baseline for detectors deployment will promote the accuracy of parameter estimation for the SMBH binary. However, the frequency for the detectable SGWB band would be lowered referring to the critical frequency,
\begin{equation}
f_\mathrm{crit} = \frac{c}{2d} = \frac{c}{ 2 \times 2 \mathrm{AU} \sin \frac{ \epsilon }{ 2 } }  \simeq \frac{0.5 \ \mathrm{mHz}}{ \sin \frac{ \epsilon }{ 2 } }
\end{equation}
where $d$ is the distance between two detectors, $\epsilon$ is the separation angle formed by the two lines connecting the Sun and detectors. 

\begin{figure}[htb]
\includegraphics[width=0.47\textwidth]{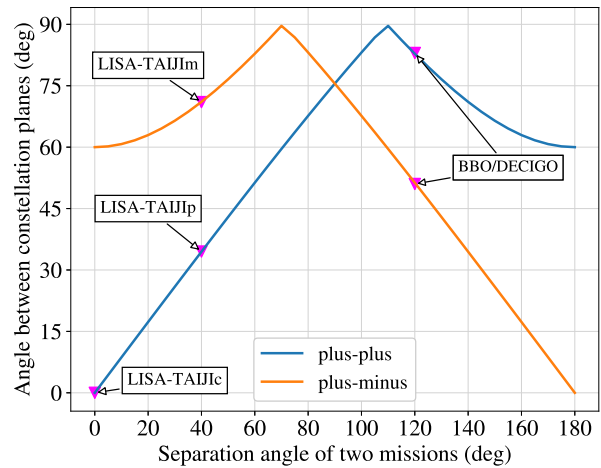} 
\caption{The angle of the formation planes varying with the separation angle of two constellations. The curve of the plus-plus indicates the two constellations having $+60^\circ$ inclinations with respect to the ecliptic plane, and the curve of the plus-minus indicates that one constellation having a $+60^\circ$ inclination and another having a $-60^\circ$ inclination.  \label{fig:angle_vs_seperation} 
}
\end{figure}

On the other hand, the angle between the constellation plane also changes with the separation angle as shown in Fig. \ref{fig:angle_vs_seperation}. Considering the orientation of the plane with an angle closing to $90^\circ$ should be helpful to resolve the source parameters, the composition of a $+60^\circ$ with a $-60^\circ$ inclination could more cooperative than the two missions with the same inclination for a separation angle small than $90^\circ$. For the BBO or DECIGO mission, the constellations are planned to be separated by $120^\circ$, and two options could be considered for their orientation deployment, $83^\circ$ or $51^\circ$ with respect to another formation plane as tagged in Fig. \ref{fig:angle_vs_seperation}.

\section{Conclusions} \label{sec:conclusions}

In this work, we investigate the performances of three alternative LISA-TAIJI networks on the sky localizations and polarization observations from the SMBH binary and the overlap reduction function for the stochastic gravitational wave background observation. For the SMBH binary system, compared to the single LISA mission, the co-located and coplanar LISA-TAIJIc network ordinarily improve the SNR and parameter resolution by a factor of $\sqrt{2}$ times. With a $1 \times 10^8$ km separation, the joint observations from LISA and TAIJIp significantly improve parameter determinations for the SMBH binary than the LISA-TAIJIc. The LISA-TAIJIm network demonstrates a better capability to determine the sky location and polarizations than the LISA-TAIJIp network as the benefits of more misaligned orientation and complementary antenna pattern.  
For the detectability for the stochastic gravitational wave background, the LISA-TAIJIc would have optimal performance as the benefit of the coplanar formation and co-location, the TAIJIm present the worst cross-correlation with the LISA due to the least aligned orientation with the LISA in the three TAIJI orbital configurations.

One lesson from this evaluation of three LISA-TAIJI networks is that the parameter resolution of the compact binary coalescences will be impacted by the SNR, the distance of detector separation, and the cooperative orientations of the detectors. 
The next generation space-based GW detectors, both DECIGO and BBO, are proposed to be the LISA-like orbit with multiple constellations to detect the relic GW left by the Big Bang, intermediate-mass black holes, etc in the deci-Hz frequency band \cite{BBO:2005,DECIGO:2006}. The parameter resolution improvements have been performed for the compact binaries in \cite{BBO:2005} as the results of the multiple interferometers and long-baseline. The orientations combinations of the constellations are also worth to be evaluated for the targeting sources.

Beyond the LISA-like orbital formation, various space missions are proposed to arrange S/C equally on a planet orbit in order to observe GW in the micro-Hz band, for instance, ASTROD-GW \cite{Ni:2012eh}, Folkner mission \cite{Baker:2019pnp}, and $\mu$Ares \cite{Sesana:2019vho}, etc. The ASTROD-GW is initially proposed to deploy 3 S/C around the Lagrange points L3, L4, and L5 of the Sun-Earth system, and an extended deployment could be 6 S/C to form two triangular interferometers to enhance the sensitivity to the SGWB \cite{Ni:2012eh}. The $\mu$Ares will place the two orthogonal triangle interferometers with respect to the Mars (or Earth/Venus) orbit. 
The trade-off of detectability from various deployments could also be explored to balance the GW observations from the compact binary systems and the cosmological stochastic background.

\begin{acknowledgments}
This work was supported by NSFC Nos. 12003059 and 11773059, the Strategic Priority Research Program of the Chinese Academy of Sciences under Grants No. XDA15021102. This work made use of the High Performance Computing Resource in the Core Facility for Advanced Research Computing at Shanghai Astronomical Observatory.
\end{acknowledgments}

%\appendix

% The \nocite command causes all entries in a bibliography to be printed out
% whether or not they are actually referenced in the text. This is appropriate
% for the sample file to show the different styles of references, but authors
% most likely will not want to use it.
\nocite{*}
\bibliography{apsref}% Produces the bibliography via BibTeX.

%apsrev4-2.bst 2019-01-14 (MD) hand-edited version of apsrev4-1.bst
%Control: key (0)
%Control: author (8) initials jnrlst
%Control: editor formatted (1) identically to author
%Control: production of article title (0) allowed
%Control: page (0) single
%Control: year (1) truncated
%Control: production of eprint (0) enabled
\providecommand{\noopsort}[1]{}\providecommand{\singleletter}[1]{#1}%
\begin{thebibliography}{71}%
\makeatletter
\providecommand \@ifxundefined [1]{%
 \@ifx{#1\undefined}
}%
\providecommand \@ifnum [1]{%
 \ifnum #1\expandafter \@firstoftwo
 \else \expandafter \@secondoftwo
 \fi
}%
\providecommand \@ifx [1]{%
 \ifx #1\expandafter \@firstoftwo
 \else \expandafter \@secondoftwo
 \fi
}%
\providecommand \natexlab [1]{#1}%
\providecommand \enquote  [1]{``#1''}%
\providecommand \bibnamefont  [1]{#1}%
\providecommand \bibfnamefont [1]{#1}%
\providecommand \citenamefont [1]{#1}%
\providecommand \href@noop [0]{\@secondoftwo}%
\providecommand \href [0]{\begingroup \@sanitize@url \@href}%
\providecommand \@href[1]{\@@startlink{#1}\@@href}%
\providecommand \@@href[1]{\endgroup#1\@@endlink}%
\providecommand \@sanitize@url [0]{\catcode `\\12\catcode `\$12\catcode
  `\&12\catcode `\#12\catcode `\^12\catcode `\_12\catcode `\%12\relax}%
\providecommand \@@startlink[1]{}%
\providecommand \@@endlink[0]{}%
\providecommand \url  [0]{\begingroup\@sanitize@url \@url }%
\providecommand \@url [1]{\endgroup\@href {#1}{\urlprefix }}%
\providecommand \urlprefix  [0]{URL }%
\providecommand \Eprint [0]{\href }%
\providecommand \doibase [0]{https://doi.org/}%
\providecommand \selectlanguage [0]{\@gobble}%
\providecommand \bibinfo  [0]{\@secondoftwo}%
\providecommand \bibfield  [0]{\@secondoftwo}%
\providecommand \translation [1]{[#1]}%
\providecommand \BibitemOpen [0]{}%
\providecommand \bibitemStop [0]{}%
\providecommand \bibitemNoStop [0]{.\EOS\space}%
\providecommand \EOS [0]{\spacefactor3000\relax}%
\providecommand \BibitemShut  [1]{\csname bibitem#1\endcsname}%
\let\auto@bib@innerbib\@empty
%</preamble>
\bibitem [{\citenamefont {Abbott}\ \emph {et~al.}(2016)\citenamefont {Abbott}
  \emph {et~al.}}]{Abbott:2016blz}%
  \BibitemOpen
  \bibfield  {author} {\bibinfo {author} {\bibfnamefont {B.~P.}\ \bibnamefont
  {Abbott}} \emph {et~al.} (\bibinfo {collaboration} {LIGO Scientific,
  Virgo}),\ }\bibfield  {title} {\bibinfo {title} {{Observation of
  Gravitational Waves from a Binary Black Hole Merger}},\ }\href
  {https://doi.org/10.1103/PhysRevLett.116.061102} {\bibfield  {journal}
  {\bibinfo  {journal} {Phys. Rev. Lett.}\ }\textbf {\bibinfo {volume} {116}},\
  \bibinfo {pages} {061102} (\bibinfo {year} {2016})},\ \Eprint
  {https://arxiv.org/abs/1602.03837} {arXiv:1602.03837 [gr-qc]} \BibitemShut
  {NoStop}%
\bibitem [{\citenamefont {Abbott}\ \emph
  {et~al.}(2017{\natexlab{a}})\citenamefont {Abbott} \emph
  {et~al.}}]{Abbott:2017oio}%
  \BibitemOpen
  \bibfield  {author} {\bibinfo {author} {\bibfnamefont {B.~P.}\ \bibnamefont
  {Abbott}} \emph {et~al.} (\bibinfo {collaboration} {LIGO Scientific,
  Virgo}),\ }\bibfield  {title} {\bibinfo {title} {{GW170814: A Three-Detector
  Observation of Gravitational Waves from a Binary Black Hole Coalescence}},\
  }\href {https://doi.org/10.1103/PhysRevLett.119.141101} {\bibfield  {journal}
  {\bibinfo  {journal} {Phys. Rev. Lett.}\ }\textbf {\bibinfo {volume} {119}},\
  \bibinfo {pages} {141101} (\bibinfo {year} {2017}{\natexlab{a}})},\ \Eprint
  {https://arxiv.org/abs/1709.09660} {arXiv:1709.09660 [gr-qc]} \BibitemShut
  {NoStop}%
\bibitem [{\citenamefont {Abbott}\ \emph
  {et~al.}(2017{\natexlab{b}})\citenamefont {Abbott} \emph
  {et~al.}}]{GW170817:detection}%
  \BibitemOpen
  \bibfield  {author} {\bibinfo {author} {\bibfnamefont {B.~P.}\ \bibnamefont
  {Abbott}} \emph {et~al.} (\bibinfo {collaboration} {LIGO Scientific,
  Virgo}),\ }\bibfield  {title} {\bibinfo {title} {{GW170817: Observation of
  Gravitational Waves from a Binary Neutron Star Inspiral}},\ }\href
  {https://doi.org/10.1103/PhysRevLett.119.161101} {\bibfield  {journal}
  {\bibinfo  {journal} {Phys. Rev. Lett.}\ }\textbf {\bibinfo {volume} {119}},\
  \bibinfo {pages} {161101} (\bibinfo {year} {2017}{\natexlab{b}})},\ \Eprint
  {https://arxiv.org/abs/1710.05832} {arXiv:1710.05832 [gr-qc]} \BibitemShut
  {NoStop}%
\bibitem [{\citenamefont {Abbott}\ \emph
  {et~al.}(2019{\natexlab{a}})\citenamefont {Abbott} \emph
  {et~al.}}]{GW170817:TGR}%
  \BibitemOpen
  \bibfield  {author} {\bibinfo {author} {\bibfnamefont {B.~P.}\ \bibnamefont
  {Abbott}} \emph {et~al.} (\bibinfo {collaboration} {LIGO Scientific,
  Virgo}),\ }\bibfield  {title} {\bibinfo {title} {{Tests of General Relativity
  with GW170817}},\ }\href {https://doi.org/10.1103/PhysRevLett.123.011102}
  {\bibfield  {journal} {\bibinfo  {journal} {Phys. Rev. Lett.}\ }\textbf
  {\bibinfo {volume} {123}},\ \bibinfo {pages} {011102} (\bibinfo {year}
  {2019}{\natexlab{a}})},\ \Eprint {https://arxiv.org/abs/1811.00364}
  {arXiv:1811.00364 [gr-qc]} \BibitemShut {NoStop}%
\bibitem [{\citenamefont {Akutsu}\ \emph {et~al.}(2020)\citenamefont {Akutsu}
  \emph {et~al.}}]{KAGRA:2020}%
  \BibitemOpen
  \bibfield  {author} {\bibinfo {author} {\bibfnamefont {T.}~\bibnamefont
  {Akutsu}} \emph {et~al.} (\bibinfo {collaboration} {KAGRA}),\ }\bibfield
  {title} {\bibinfo {title} {{Overview of KAGRA : KAGRA science}}\ }\href
  {https://doi.org/10.1093/ptep/ptaa120} {10.1093/ptep/ptaa120} (\bibinfo
  {year} {2020}),\ \Eprint {https://arxiv.org/abs/2008.02921} {arXiv:2008.02921
  [gr-qc]} \BibitemShut {NoStop}%
\bibitem [{\citenamefont {Abbott}\ \emph {et~al.}(2020)\citenamefont {Abbott}
  \emph {et~al.}}]{LVK:LLR}%
  \BibitemOpen
  \bibfield  {author} {\bibinfo {author} {\bibfnamefont {B.~P.}\ \bibnamefont
  {Abbott}} \emph {et~al.} (\bibinfo {collaboration} {KAGRA, LIGO Scientific,
  Virgo}),\ }\bibfield  {title} {\bibinfo {title} {{Prospects for observing and
  localizing gravitational-wave transients with Advanced LIGO, Advanced Virgo
  and KAGRA}},\ }\href {https://doi.org/10.1007/s41114-020-00026-9} {\bibfield
  {journal} {\bibinfo  {journal} {Living Rev. Rel.}\ }\textbf {\bibinfo
  {volume} {23}},\ \bibinfo {pages} {3} (\bibinfo {year} {2020})}\BibitemShut
  {NoStop}%
\bibitem [{\citenamefont {Schutz}(2011)}]{Schutz:2011tw}%
  \BibitemOpen
  \bibfield  {author} {\bibinfo {author} {\bibfnamefont {B.~F.}\ \bibnamefont
  {Schutz}},\ }\bibfield  {title} {\bibinfo {title} {{Networks of gravitational
  wave detectors and three figures of merit}},\ }\href
  {https://doi.org/10.1088/0264-9381/28/12/125023} {\bibfield  {journal}
  {\bibinfo  {journal} {Class. Quant. Grav.}\ }\textbf {\bibinfo {volume}
  {28}},\ \bibinfo {pages} {125023} (\bibinfo {year} {2011})},\ \Eprint
  {https://arxiv.org/abs/1102.5421} {arXiv:1102.5421 [astro-ph.IM]}
  \BibitemShut {NoStop}%
\bibitem [{\citenamefont {Abbott}\ \emph
  {et~al.}(2017{\natexlab{c}})\citenamefont {Abbott} \emph
  {et~al.}}]{TheLIGOScientific:2016dpb}%
  \BibitemOpen
  \bibfield  {author} {\bibinfo {author} {\bibfnamefont {B.~P.}\ \bibnamefont
  {Abbott}} \emph {et~al.} (\bibinfo {collaboration} {LIGO Scientific,
  Virgo}),\ }\bibfield  {title} {\bibinfo {title} {{Upper Limits on the
  Stochastic Gravitational-Wave Background from Advanced LIGO\textquoteright{}s
  First Observing Run}},\ }\href
  {https://doi.org/10.1103/PhysRevLett.118.121101} {\bibfield  {journal}
  {\bibinfo  {journal} {Phys. Rev. Lett.}\ }\textbf {\bibinfo {volume} {118}},\
  \bibinfo {pages} {121101} (\bibinfo {year} {2017}{\natexlab{c}})},\ \bibinfo
  {note} {[Erratum: Phys.Rev.Lett. 119, 029901 (2017)]},\ \Eprint
  {https://arxiv.org/abs/1612.02029} {arXiv:1612.02029 [gr-qc]} \BibitemShut
  {NoStop}%
\bibitem [{\citenamefont {Abbott}\ \emph {et~al.}(2018)\citenamefont {Abbott}
  \emph {et~al.}}]{Abbott:2018utx}%
  \BibitemOpen
  \bibfield  {author} {\bibinfo {author} {\bibfnamefont {B.~P.}\ \bibnamefont
  {Abbott}} \emph {et~al.} (\bibinfo {collaboration} {LIGO Scientific,
  Virgo}),\ }\bibfield  {title} {\bibinfo {title} {{Search for Tensor, Vector,
  and Scalar Polarizations in the Stochastic Gravitational-Wave Background}},\
  }\href {https://doi.org/10.1103/PhysRevLett.120.201102} {\bibfield  {journal}
  {\bibinfo  {journal} {Phys. Rev. Lett.}\ }\textbf {\bibinfo {volume} {120}},\
  \bibinfo {pages} {201102} (\bibinfo {year} {2018})},\ \Eprint
  {https://arxiv.org/abs/1802.10194} {arXiv:1802.10194 [gr-qc]} \BibitemShut
  {NoStop}%
\bibitem [{\citenamefont {Abbott}\ \emph
  {et~al.}(2019{\natexlab{b}})\citenamefont {Abbott} \emph
  {et~al.}}]{LIGOScientific:2019vic}%
  \BibitemOpen
  \bibfield  {author} {\bibinfo {author} {\bibfnamefont {B.~P.}\ \bibnamefont
  {Abbott}} \emph {et~al.} (\bibinfo {collaboration} {LIGO Scientific,
  Virgo}),\ }\bibfield  {title} {\bibinfo {title} {{Search for the isotropic
  stochastic background using data from Advanced LIGO\textquoteright{}s second
  observing run}},\ }\href {https://doi.org/10.1103/PhysRevD.100.061101}
  {\bibfield  {journal} {\bibinfo  {journal} {Phys. Rev. D}\ }\textbf {\bibinfo
  {volume} {100}},\ \bibinfo {pages} {061101} (\bibinfo {year}
  {2019}{\natexlab{b}})},\ \Eprint {https://arxiv.org/abs/1903.02886}
  {arXiv:1903.02886 [gr-qc]} \BibitemShut {NoStop}%
\bibitem [{\citenamefont {Abbott}\ \emph
  {et~al.}(2021{\natexlab{a}})\citenamefont {Abbott} \emph
  {et~al.}}]{Abbott:2021xxi}%
  \BibitemOpen
  \bibfield  {author} {\bibinfo {author} {\bibfnamefont {R.}~\bibnamefont
  {Abbott}} \emph {et~al.} (\bibinfo {collaboration} {LIGO Scientific, Virgo,
  KAGRA}),\ }\bibfield  {title} {\bibinfo {title} {{Upper Limits on the
  Isotropic Gravitational-Wave Background from Advanced LIGO's and Advanced
  Virgo's Third Observing Run}},\ }\href@noop {} {\  (\bibinfo {year}
  {2021}{\natexlab{a}})},\ \Eprint {https://arxiv.org/abs/2101.12130}
  {arXiv:2101.12130 [gr-qc]} \BibitemShut {NoStop}%
\bibitem [{\citenamefont {Abbott}\ \emph
  {et~al.}(2021{\natexlab{b}})\citenamefont {Abbott} \emph
  {et~al.}}]{Abbott:2021ksc}%
  \BibitemOpen
  \bibfield  {author} {\bibinfo {author} {\bibfnamefont {R.}~\bibnamefont
  {Abbott}} \emph {et~al.} (\bibinfo {collaboration} {LIGO Scientific, Virgo,
  KAGRA}),\ }\bibfield  {title} {\bibinfo {title} {{Constraints on cosmic
  strings using data from the third Advanced LIGO-Virgo observing run}},\
  }\href@noop {} {\  (\bibinfo {year} {2021}{\natexlab{b}})},\ \Eprint
  {https://arxiv.org/abs/2101.12248} {arXiv:2101.12248 [gr-qc]} \BibitemShut
  {NoStop}%
\bibitem [{\citenamefont {Abbott}\ \emph
  {et~al.}(2021{\natexlab{c}})\citenamefont {Abbott} \emph
  {et~al.}}]{Abbott:2021jel}%
  \BibitemOpen
  \bibfield  {author} {\bibinfo {author} {\bibfnamefont {R.}~\bibnamefont
  {Abbott}} \emph {et~al.} (\bibinfo {collaboration} {LIGO Scientific, Virgo,
  KAGRA}),\ }\bibfield  {title} {\bibinfo {title} {{Search for anisotropic
  gravitational-wave backgrounds using data from Advanced LIGO's and Advanced
  Virgo's first three observing runs}},\ }\href@noop {} {\  (\bibinfo {year}
  {2021}{\natexlab{c}})},\ \Eprint {https://arxiv.org/abs/2103.08520}
  {arXiv:2103.08520 [gr-qc]} \BibitemShut {NoStop}%
\bibitem [{\citenamefont {Crowder}\ and\ \citenamefont
  {Cornish}(2005)}]{BBO:2005}%
  \BibitemOpen
  \bibfield  {author} {\bibinfo {author} {\bibfnamefont {J.}~\bibnamefont
  {Crowder}}\ and\ \bibinfo {author} {\bibfnamefont {N.~J.}\ \bibnamefont
  {Cornish}},\ }\bibfield  {title} {\bibinfo {title} {{Beyond LISA: Exploring
  future gravitational wave missions}},\ }\href
  {https://doi.org/10.1103/PhysRevD.72.083005} {\bibfield  {journal} {\bibinfo
  {journal} {Phys. Rev. D}\ }\textbf {\bibinfo {volume} {72}},\ \bibinfo
  {pages} {083005} (\bibinfo {year} {2005})},\ \Eprint
  {https://arxiv.org/abs/gr-qc/0506015} {arXiv:gr-qc/0506015} \BibitemShut
  {NoStop}%
\bibitem [{\citenamefont {Kawamura}\ \emph {et~al.}(2006)\citenamefont
  {Kawamura} \emph {et~al.}}]{DECIGO:2006}%
  \BibitemOpen
  \bibfield  {author} {\bibinfo {author} {\bibfnamefont {S.}~\bibnamefont
  {Kawamura}} \emph {et~al.},\ }\bibfield  {title} {\bibinfo {title} {{The
  Japanese space gravitational wave antenna DECIGO}},\ }\href
  {https://doi.org/10.1088/0264-9381/23/8/S17} {\bibfield  {journal} {\bibinfo
  {journal} {Class. Quant. Grav.}\ }\textbf {\bibinfo {volume} {23}},\ \bibinfo
  {pages} {S125} (\bibinfo {year} {2006})}\BibitemShut {NoStop}%
\bibitem [{\citenamefont {Romano}\ and\ \citenamefont
  {Cornish}(2017)}]{Romano:2016dpx}%
  \BibitemOpen
  \bibfield  {author} {\bibinfo {author} {\bibfnamefont {J.~D.}\ \bibnamefont
  {Romano}}\ and\ \bibinfo {author} {\bibfnamefont {N.~J.}\ \bibnamefont
  {Cornish}},\ }\bibfield  {title} {\bibinfo {title} {{Detection methods for
  stochastic gravitational-wave backgrounds: a unified treatment}},\ }\href
  {https://doi.org/10.1007/s41114-017-0004-1} {\bibfield  {journal} {\bibinfo
  {journal} {Living Rev. Rel.}\ }\textbf {\bibinfo {volume} {20}},\ \bibinfo
  {pages} {2} (\bibinfo {year} {2017})},\ \Eprint
  {https://arxiv.org/abs/1608.06889} {arXiv:1608.06889 [gr-qc]} \BibitemShut
  {NoStop}%
\bibitem [{\citenamefont {Schmitz}(2021)}]{Schmitz:2020syl}%
  \BibitemOpen
  \bibfield  {author} {\bibinfo {author} {\bibfnamefont {K.}~\bibnamefont
  {Schmitz}},\ }\bibfield  {title} {\bibinfo {title} {{New Sensitivity Curves
  for Gravitational-Wave Signals from Cosmological Phase Transitions}},\ }\href
  {https://doi.org/10.1007/JHEP01(2021)097} {\bibfield  {journal} {\bibinfo
  {journal} {JHEP}\ }\textbf {\bibinfo {volume} {01}},\ \bibinfo {pages}
  {097}},\ \Eprint {https://arxiv.org/abs/2002.04615} {arXiv:2002.04615
  [hep-ph]} \BibitemShut {NoStop}%
\bibitem [{\citenamefont {Ni}(2020)}]{Ni:2020utm}%
  \BibitemOpen
  \bibfield  {author} {\bibinfo {author} {\bibfnamefont {W.-T.}\ \bibnamefont
  {Ni}},\ }\bibfield  {title} {\bibinfo {title} {{Mid-frequency gravitational
  wave detection and sources}},\ }\href
  {https://doi.org/10.1142/S021827181902005X} {\bibfield  {journal} {\bibinfo
  {journal} {Int. J. Mod. Phys. D}\ }\textbf {\bibinfo {volume} {29}},\
  \bibinfo {pages} {1902005} (\bibinfo {year} {2020})},\ \Eprint
  {https://arxiv.org/abs/2004.05590} {arXiv:2004.05590 [gr-qc]} \BibitemShut
  {NoStop}%
\bibitem [{\citenamefont {{Amaro-Seoane}}\ \emph {et~al.}(2017)\citenamefont
  {{Amaro-Seoane}}, \citenamefont {{Audley}}, \citenamefont {{Babak}},\ and\
  \citenamefont {{et al}}}]{2017arXiv170200786A}%
  \BibitemOpen
  \bibfield  {author} {\bibinfo {author} {\bibfnamefont {P.}~\bibnamefont
  {{Amaro-Seoane}}}, \bibinfo {author} {\bibfnamefont {H.}~\bibnamefont
  {{Audley}}}, \bibinfo {author} {\bibfnamefont {S.}~\bibnamefont {{Babak}}},\
  and\ \bibinfo {author} {\bibnamefont {{et al}}} (\bibinfo {collaboration}
  {{LISA Team}}),\ }\bibfield  {title} {\bibinfo {title} {{Laser Interferometer
  Space Antenna}},\ }\href@noop {} {\bibfield  {journal} {\bibinfo  {journal}
  {arXiv e-prints}\ ,\ \bibinfo {eid} {arXiv:1702.00786}} (\bibinfo {year}
  {2017})}\BibitemShut {NoStop}%
\bibitem [{\citenamefont {Hu}\ and\ \citenamefont {Wu}(2017)}]{Hu:2017mde}%
  \BibitemOpen
  \bibfield  {author} {\bibinfo {author} {\bibfnamefont {W.-R.}\ \bibnamefont
  {Hu}}\ and\ \bibinfo {author} {\bibfnamefont {Y.-L.}\ \bibnamefont {Wu}},\
  }\bibfield  {title} {\bibinfo {title} {{The Taiji Program in Space for
  gravitational wave physics and the nature of gravity}},\ }\href
  {https://doi.org/10.1093/nsr/nwx116} {\bibfield  {journal} {\bibinfo
  {journal} {Natl. Sci. Rev.}\ }\textbf {\bibinfo {volume} {4}},\ \bibinfo
  {pages} {685} (\bibinfo {year} {2017})}\BibitemShut {NoStop}%
%%CITATION = INSPIRE-1712806;%%
\bibitem [{\citenamefont {Luo}\ \emph {et~al.}(2016)\citenamefont {Luo} \emph
  {et~al.}}]{Luo:2015ght}%
  \BibitemOpen
  \bibfield  {author} {\bibinfo {author} {\bibfnamefont {J.}~\bibnamefont
  {Luo}} \emph {et~al.} (\bibinfo {collaboration} {TianQin Team}),\ }\bibfield
  {title} {\bibinfo {title} {{TianQin: a space-borne gravitational wave
  detector}},\ }\href {https://doi.org/10.1088/0264-9381/33/3/035010}
  {\bibfield  {journal} {\bibinfo  {journal} {Class. Quant. Grav.}\ }\textbf
  {\bibinfo {volume} {33}},\ \bibinfo {pages} {035010} (\bibinfo {year}
  {2016})},\ \Eprint {https://arxiv.org/abs/1512.02076} {arXiv:1512.02076
  [astro-ph.IM]} \BibitemShut {NoStop}%
%%CITATION = ARXIV:1512.02076;%%
\bibitem [{\citenamefont {Dhurandhar}\ \emph {et~al.}(2005)\citenamefont
  {Dhurandhar}, \citenamefont {Rajesh~Nayak}, \citenamefont {Koshti},\ and\
  \citenamefont {Vinet}}]{Dhurandhar:2004rv}%
  \BibitemOpen
  \bibfield  {author} {\bibinfo {author} {\bibfnamefont {S.~V.}\ \bibnamefont
  {Dhurandhar}}, \bibinfo {author} {\bibfnamefont {K.}~\bibnamefont
  {Rajesh~Nayak}}, \bibinfo {author} {\bibfnamefont {S.}~\bibnamefont
  {Koshti}},\ and\ \bibinfo {author} {\bibfnamefont {J.~Y.}\ \bibnamefont
  {Vinet}},\ }\bibfield  {title} {\bibinfo {title} {{Fundamentals of the LISA
  stable flight formation}},\ }\href
  {https://doi.org/10.1088/0264-9381/22/3/002} {\bibfield  {journal} {\bibinfo
  {journal} {Class. Quant. Grav.}\ }\textbf {\bibinfo {volume} {22}},\ \bibinfo
  {pages} {481} (\bibinfo {year} {2005})},\ \Eprint
  {https://arxiv.org/abs/gr-qc/0410093} {arXiv:gr-qc/0410093} \BibitemShut
  {NoStop}%
\bibitem [{\citenamefont {Ruan}\ \emph {et~al.}(2020)\citenamefont {Ruan},
  \citenamefont {Liu}, \citenamefont {Guo}, \citenamefont {Wu},\ and\
  \citenamefont {Cai}}]{Ruan:2020smc}%
  \BibitemOpen
  \bibfield  {author} {\bibinfo {author} {\bibfnamefont {W.-H.}\ \bibnamefont
  {Ruan}}, \bibinfo {author} {\bibfnamefont {C.}~\bibnamefont {Liu}}, \bibinfo
  {author} {\bibfnamefont {Z.-K.}\ \bibnamefont {Guo}}, \bibinfo {author}
  {\bibfnamefont {Y.-L.}\ \bibnamefont {Wu}},\ and\ \bibinfo {author}
  {\bibfnamefont {R.-G.}\ \bibnamefont {Cai}},\ }\bibfield  {title} {\bibinfo
  {title} {{The LISA-Taiji network}},\ }\href
  {https://doi.org/10.1038/s41550-019-1008-4} {\bibfield  {journal} {\bibinfo
  {journal} {Nature Astron.}\ }\textbf {\bibinfo {volume} {4}},\ \bibinfo
  {pages} {108} (\bibinfo {year} {2020})},\ \Eprint
  {https://arxiv.org/abs/2002.03603} {arXiv:2002.03603 [gr-qc]} \BibitemShut
  {NoStop}%
\bibitem [{\citenamefont {{Wang}}\ \emph {et~al.}(2020)\citenamefont {{Wang}},
  \citenamefont {{Ni}}, \citenamefont {{Han}}, \citenamefont {{Yang}},\ and\
  \citenamefont {{Zhong}}}]{Wang:2020a}%
  \BibitemOpen
  \bibfield  {author} {\bibinfo {author} {\bibfnamefont {G.}~\bibnamefont
  {{Wang}}}, \bibinfo {author} {\bibfnamefont {W.-T.}\ \bibnamefont {{Ni}}},
  \bibinfo {author} {\bibfnamefont {W.-B.}\ \bibnamefont {{Han}}}, \bibinfo
  {author} {\bibfnamefont {S.-C.}\ \bibnamefont {{Yang}}},\ and\ \bibinfo
  {author} {\bibfnamefont {X.-Y.}\ \bibnamefont {{Zhong}}},\ }\bibfield
  {title} {\bibinfo {title} {{Numerical simulation of sky localization for
  LISA-TAIJI joint observation}},\ }\href
  {https://doi.org/10.1103/PhysRevD.102.024089} {\bibfield  {journal} {\bibinfo
   {journal} {\prd}\ }\textbf {\bibinfo {volume} {102}},\ \bibinfo {pages}
  {024089} (\bibinfo {year} {2020})},\ \Eprint
  {https://arxiv.org/abs/2002.12628} {arXiv:2002.12628} \BibitemShut {NoStop}%
\bibitem [{\citenamefont {Omiya}\ and\ \citenamefont
  {Seto}(2020)}]{Omiya:2020fvw}%
  \BibitemOpen
  \bibfield  {author} {\bibinfo {author} {\bibfnamefont {H.}~\bibnamefont
  {Omiya}}\ and\ \bibinfo {author} {\bibfnamefont {N.}~\bibnamefont {Seto}},\
  }\bibfield  {title} {\bibinfo {title} {{Searching for anomalous polarization
  modes of the stochastic gravitational wave background with LISA and Taiji}},\
  }\href {https://doi.org/10.1103/PhysRevD.102.084053} {\bibfield  {journal}
  {\bibinfo  {journal} {Phys. Rev. D}\ }\textbf {\bibinfo {volume} {102}},\
  \bibinfo {pages} {084053} (\bibinfo {year} {2020})},\ \Eprint
  {https://arxiv.org/abs/2010.00771} {arXiv:2010.00771 [gr-qc]} \BibitemShut
  {NoStop}%
\bibitem [{\citenamefont {Seto}(2020)}]{Seto:2020mfd}%
  \BibitemOpen
  \bibfield  {author} {\bibinfo {author} {\bibfnamefont {N.}~\bibnamefont
  {Seto}},\ }\bibfield  {title} {\bibinfo {title} {{Gravitational Wave
  Background Search by Correlating Multiple Triangular Detectors in the mHz
  Band}},\ }\href {https://doi.org/10.1103/PhysRevD.102.123547} {\bibfield
  {journal} {\bibinfo  {journal} {Phys. Rev. D}\ }\textbf {\bibinfo {volume}
  {102}},\ \bibinfo {pages} {123547} (\bibinfo {year} {2020})},\ \Eprint
  {https://arxiv.org/abs/2010.06877} {arXiv:2010.06877 [gr-qc]} \BibitemShut
  {NoStop}%
\bibitem [{\citenamefont {Orlando}\ \emph {et~al.}(2020)\citenamefont
  {Orlando}, \citenamefont {Pieroni},\ and\ \citenamefont
  {Ricciardone}}]{Orlando:2020oko}%
  \BibitemOpen
  \bibfield  {author} {\bibinfo {author} {\bibfnamefont {G.}~\bibnamefont
  {Orlando}}, \bibinfo {author} {\bibfnamefont {M.}~\bibnamefont {Pieroni}},\
  and\ \bibinfo {author} {\bibfnamefont {A.}~\bibnamefont {Ricciardone}},\
  }\bibfield  {title} {\bibinfo {title} {{Measuring Parity Violation in the
  Stochastic Gravitational Wave Background with the LISA-Taiji network}},\
  }\href@noop {} {\  (\bibinfo {year} {2020})},\ \Eprint
  {https://arxiv.org/abs/2011.07059} {arXiv:2011.07059 [astro-ph.CO]}
  \BibitemShut {NoStop}%
\bibitem [{\citenamefont {Wang}\ \emph
  {et~al.}(2020{\natexlab{a}})\citenamefont {Wang}, \citenamefont {Ruan},
  \citenamefont {Yang}, \citenamefont {Guo}, \citenamefont {Cai},\ and\
  \citenamefont {Hu}}]{Wang:2020dkc}%
  \BibitemOpen
  \bibfield  {author} {\bibinfo {author} {\bibfnamefont {R.}~\bibnamefont
  {Wang}}, \bibinfo {author} {\bibfnamefont {W.-H.}\ \bibnamefont {Ruan}},
  \bibinfo {author} {\bibfnamefont {Q.}~\bibnamefont {Yang}}, \bibinfo {author}
  {\bibfnamefont {Z.-K.}\ \bibnamefont {Guo}}, \bibinfo {author} {\bibfnamefont
  {R.-G.}\ \bibnamefont {Cai}},\ and\ \bibinfo {author} {\bibfnamefont
  {B.}~\bibnamefont {Hu}},\ }\bibfield  {title} {\bibinfo {title} {{Hubble
  parameter estimation via dark sirens with the LISA-Taiji network}},\
  }\href@noop {} {\  (\bibinfo {year} {2020}{\natexlab{a}})},\ \Eprint
  {https://arxiv.org/abs/2010.14732} {arXiv:2010.14732 [astro-ph.CO]}
  \BibitemShut {NoStop}%
\bibitem [{\citenamefont {Wang}\ \emph {et~al.}(2021)\citenamefont {Wang},
  \citenamefont {Jin}, \citenamefont {Zhang},\ and\ \citenamefont
  {Zhang}}]{Wang:2021srv}%
  \BibitemOpen
  \bibfield  {author} {\bibinfo {author} {\bibfnamefont {L.-F.}\ \bibnamefont
  {Wang}}, \bibinfo {author} {\bibfnamefont {S.-J.}\ \bibnamefont {Jin}},
  \bibinfo {author} {\bibfnamefont {J.-F.}\ \bibnamefont {Zhang}},\ and\
  \bibinfo {author} {\bibfnamefont {X.}~\bibnamefont {Zhang}},\ }\bibfield
  {title} {\bibinfo {title} {{Forecast for cosmological parameter estimation
  with gravitational-wave standard sirens from the LISA-Taiji network}},\
  }\href@noop {} {\  (\bibinfo {year} {2021})},\ \Eprint
  {https://arxiv.org/abs/2101.11882} {arXiv:2101.11882 [gr-qc]} \BibitemShut
  {NoStop}%
\bibitem [{\citenamefont {Wang}\ and\ \citenamefont
  {Han}(2021)}]{Wang:2021polar}%
  \BibitemOpen
  \bibfield  {author} {\bibinfo {author} {\bibfnamefont {G.}~\bibnamefont
  {Wang}}\ and\ \bibinfo {author} {\bibfnamefont {W.-B.}\ \bibnamefont {Han}},\
  }\bibfield  {title} {\bibinfo {title} {{Observing gravitational wave
  polarizations with the LISA-TAIJI network}},\ }\href
  {https://doi.org/10.1103/PhysRevD.103.064021} {\bibfield  {journal} {\bibinfo
   {journal} {Phys. Rev. D}\ }\textbf {\bibinfo {volume} {103}},\ \bibinfo
  {pages} {064021} (\bibinfo {year} {2021})},\ \Eprint
  {https://arxiv.org/abs/2101.01991} {arXiv:2101.01991 [gr-qc]} \BibitemShut
  {NoStop}%
\bibitem [{\citenamefont {{LISA~Study~Team}}(2000)}]{LISA2000}%
  \BibitemOpen
  \bibfield  {author} {\bibinfo {author} {\bibnamefont {{LISA~Study~Team}}},\
  }\href@noop {} {\emph {\bibinfo {title} {LISA (Laser Interferometer Space
  Antenna): A Cornerstone Mission for the Observation of Gravitational
  Waves}}},\ \bibinfo {type} {Tech. Rep.}\ \bibinfo {number} {11}\ (\bibinfo
  {institution} {ESA-SCI},\ \bibinfo {year} {2000})\ \bibinfo {note} {system
  and Technology Study Report}\BibitemShut {NoStop}%
\bibitem [{\citenamefont {Wang}\ and\ \citenamefont {Ni}(2019)}]{Wang:2017aqq}%
  \BibitemOpen
  \bibfield  {author} {\bibinfo {author} {\bibfnamefont {G.}~\bibnamefont
  {Wang}}\ and\ \bibinfo {author} {\bibfnamefont {W.-T.}\ \bibnamefont {Ni}},\
  }\bibfield  {title} {\bibinfo {title} {{Numerical simulation of time delay
  interferometry for TAIJI and new LISA}},\ }\href
  {https://doi.org/10.1088/1674-4527/19/4/58} {\bibfield  {journal} {\bibinfo
  {journal} {Res. Astron. Astrophys.}\ }\textbf {\bibinfo {volume} {19}},\
  \bibinfo {pages} {058} (\bibinfo {year} {2019})},\ \Eprint
  {https://arxiv.org/abs/1707.09127} {arXiv:1707.09127 [astro-ph.IM]}
  \BibitemShut {NoStop}%
%%CITATION = ARXIV:1707.09127;%%
\bibitem [{\citenamefont {Wang}\ and\ \citenamefont {Ni}(2012)}]{Wang:2012aea}%
  \BibitemOpen
  \bibfield  {author} {\bibinfo {author} {\bibfnamefont {G.}~\bibnamefont
  {Wang}}\ and\ \bibinfo {author} {\bibfnamefont {W.-T.}\ \bibnamefont {Ni}},\
  }\bibfield  {title} {\bibinfo {title} {{Time-delay Interferometry for
  ASTROD-GW}},\ }\href {https://doi.org/10.1016/j.chinastron.2012.04.009}
  {\bibfield  {journal} {\bibinfo  {journal} {Chin. Astron. Astrophys.}\
  }\textbf {\bibinfo {volume} {36}},\ \bibinfo {pages} {211} (\bibinfo {year}
  {2012})},\ \bibinfo {note} {and references therein}\BibitemShut {NoStop}%
%%CITATION = CASGE,36,211;%%
\bibitem [{\citenamefont {Wang}\ and\ \citenamefont
  {Ni}(2013{\natexlab{a}})}]{Wang:2012ce}%
  \BibitemOpen
  \bibfield  {author} {\bibinfo {author} {\bibfnamefont {G.}~\bibnamefont
  {Wang}}\ and\ \bibinfo {author} {\bibfnamefont {W.-T.}\ \bibnamefont {Ni}},\
  }\bibfield  {title} {\bibinfo {title} {{Numermcal simulation of time delay
  interferometry for NGO/eLISA}},\ }\href
  {https://doi.org/10.1088/0264-9381/30/6/065011} {\bibfield  {journal}
  {\bibinfo  {journal} {Class. Quant. Grav.}\ }\textbf {\bibinfo {volume}
  {30}},\ \bibinfo {pages} {065011} (\bibinfo {year} {2013}{\natexlab{a}})},\
  \Eprint {https://arxiv.org/abs/1204.2125} {arXiv:1204.2125 [gr-qc]}
  \BibitemShut {NoStop}%
%%CITATION = ARXIV:1204.2125;%%
\bibitem [{\citenamefont {Wang}\ and\ \citenamefont
  {Ni}(2013{\natexlab{b}})}]{Wang:2013te}%
  \BibitemOpen
  \bibfield  {author} {\bibinfo {author} {\bibfnamefont {G.}~\bibnamefont
  {Wang}}\ and\ \bibinfo {author} {\bibfnamefont {W.-T.}\ \bibnamefont {Ni}},\
  }\bibfield  {title} {\bibinfo {title} {{Orbit optimization for ASTROD-GW and
  its time delay interferometry with two arms using CGC ephemeris}},\ }\href
  {https://doi.org/10.1088/1674-1056/22/4/049501} {\bibfield  {journal}
  {\bibinfo  {journal} {Chin. Phys.}\ }\textbf {\bibinfo {volume} {B22}},\
  \bibinfo {pages} {049501} (\bibinfo {year} {2013}{\natexlab{b}})},\ \Eprint
  {https://arxiv.org/abs/1205.5175} {arXiv:1205.5175 [gr-qc]} \BibitemShut
  {NoStop}%
%%CITATION = ARXIV:1205.5175;%%
\bibitem [{\citenamefont {Dhurandhar}\ \emph {et~al.}(2013)\citenamefont
  {Dhurandhar}, \citenamefont {Ni},\ and\ \citenamefont
  {Wang}}]{Dhurandhar:2011ik}%
  \BibitemOpen
  \bibfield  {author} {\bibinfo {author} {\bibfnamefont {S.~V.}\ \bibnamefont
  {Dhurandhar}}, \bibinfo {author} {\bibfnamefont {W.~T.}\ \bibnamefont {Ni}},\
  and\ \bibinfo {author} {\bibfnamefont {G.}~\bibnamefont {Wang}},\ }\bibfield
  {title} {\bibinfo {title} {{Numerical simulation of time delay interferometry
  for a LISA-like mission with the simplification of having only one
  interferometer}},\ }\href {https://doi.org/10.1016/j.asr.2012.09.017}
  {\bibfield  {journal} {\bibinfo  {journal} {Adv. Space Res.}\ }\textbf
  {\bibinfo {volume} {51}},\ \bibinfo {pages} {198} (\bibinfo {year} {2013})},\
  \Eprint {https://arxiv.org/abs/1102.4965} {arXiv:1102.4965 [gr-qc]}
  \BibitemShut {NoStop}%
%%CITATION = ARXIV:1102.4965;%%
\bibitem [{\citenamefont {Wang}\ and\ \citenamefont {Ni}(2015)}]{Wang:2014cla}%
  \BibitemOpen
  \bibfield  {author} {\bibinfo {author} {\bibfnamefont {G.}~\bibnamefont
  {Wang}}\ and\ \bibinfo {author} {\bibfnamefont {W.-T.}\ \bibnamefont {Ni}},\
  }\bibfield  {title} {\bibinfo {title} {{Orbit optimization and time delay
  interferometry for inclined ASTROD-GW formation with half-year
  precession-period}},\ }\href {https://doi.org/10.1088/1674-1056/24/5/059501}
  {\bibfield  {journal} {\bibinfo  {journal} {Chin. Phys.}\ }\textbf {\bibinfo
  {volume} {B24}},\ \bibinfo {pages} {059501} (\bibinfo {year} {2015})},\
  \Eprint {https://arxiv.org/abs/1409.4162} {arXiv:1409.4162 [gr-qc]}
  \BibitemShut {NoStop}%
%%CITATION = ARXIV:1409.4162;%%
\bibitem [{\citenamefont {Wang}\ \emph
  {et~al.}(2020{\natexlab{b}})\citenamefont {Wang}, \citenamefont {Ni},\ and\
  \citenamefont {Han}}]{Wang:1stTDI}%
  \BibitemOpen
  \bibfield  {author} {\bibinfo {author} {\bibfnamefont {G.}~\bibnamefont
  {Wang}}, \bibinfo {author} {\bibfnamefont {W.-T.}\ \bibnamefont {Ni}},\ and\
  \bibinfo {author} {\bibfnamefont {W.-B.}\ \bibnamefont {Han}},\ }\bibfield
  {title} {\bibinfo {title} {{Revisiting time delay interferometry for
  unequal-arm LISA and TAIJI}},\ }\href@noop {} {\  (\bibinfo {year}
  {2020}{\natexlab{b}})},\ \Eprint {https://arxiv.org/abs/2008.05812}
  {arXiv:2008.05812 [gr-qc]} \BibitemShut {NoStop}%
\bibitem [{\citenamefont {Wang}\ \emph
  {et~al.}(2020{\natexlab{c}})\citenamefont {Wang}, \citenamefont {Ni},
  \citenamefont {Han},\ and\ \citenamefont {Qiao}}]{Wang:2ndTDI}%
  \BibitemOpen
  \bibfield  {author} {\bibinfo {author} {\bibfnamefont {G.}~\bibnamefont
  {Wang}}, \bibinfo {author} {\bibfnamefont {W.-T.}\ \bibnamefont {Ni}},
  \bibinfo {author} {\bibfnamefont {W.-B.}\ \bibnamefont {Han}},\ and\ \bibinfo
  {author} {\bibfnamefont {C.-F.}\ \bibnamefont {Qiao}},\ }\bibfield  {title}
  {\bibinfo {title} {{Algorithm for TDI numerical simulation and sensitivity
  investigation}},\ }\href@noop {} {\  (\bibinfo {year}
  {2020}{\natexlab{c}})},\ \Eprint {https://arxiv.org/abs/2010.15544}
  {arXiv:2010.15544 [gr-qc]} \BibitemShut {NoStop}%
\bibitem [{\citenamefont {{Estabrook}}\ and\ \citenamefont
  {{Wahlquist}}(1975)}]{1975GReGr...6..439E}%
  \BibitemOpen
  \bibfield  {author} {\bibinfo {author} {\bibfnamefont {F.~B.}\ \bibnamefont
  {{Estabrook}}}\ and\ \bibinfo {author} {\bibfnamefont {H.~D.}\ \bibnamefont
  {{Wahlquist}}},\ }\bibfield  {title} {\bibinfo {title} {{Response of Doppler
  spacecraft tracking to gravitational radiation.}},\ }\href
  {https://doi.org/10.1007/BF00762449} {\bibfield  {journal} {\bibinfo
  {journal} {General Relativity and Gravitation}\ }\textbf {\bibinfo {volume}
  {6}},\ \bibinfo {pages} {439} (\bibinfo {year} {1975})}\BibitemShut {NoStop}%
\bibitem [{\citenamefont {{Wahlquist}}(1987)}]{1987GReGr..19.1101W}%
  \BibitemOpen
  \bibfield  {author} {\bibinfo {author} {\bibfnamefont {H.}~\bibnamefont
  {{Wahlquist}}},\ }\bibfield  {title} {\bibinfo {title} {{The Doppler response
  to gravitational waves from a binary star source.}},\ }\href
  {https://doi.org/10.1007/BF00759146} {\bibfield  {journal} {\bibinfo
  {journal} {General Relativity and Gravitation}\ }\textbf {\bibinfo {volume}
  {19}},\ \bibinfo {pages} {1101} (\bibinfo {year} {1987})}\BibitemShut
  {NoStop}%
\bibitem [{\citenamefont {Vallisneri}\ \emph {et~al.}(2008)\citenamefont
  {Vallisneri}, \citenamefont {Crowder},\ and\ \citenamefont
  {Tinto}}]{Vallisneri:2007xa}%
  \BibitemOpen
  \bibfield  {author} {\bibinfo {author} {\bibfnamefont {M.}~\bibnamefont
  {Vallisneri}}, \bibinfo {author} {\bibfnamefont {J.}~\bibnamefont
  {Crowder}},\ and\ \bibinfo {author} {\bibfnamefont {M.}~\bibnamefont
  {Tinto}},\ }\bibfield  {title} {\bibinfo {title} {{Sensitivity and
  parameter-estimation precision for alternate LISA configurations}},\ }\href
  {https://doi.org/10.1088/0264-9381/25/6/065005} {\bibfield  {journal}
  {\bibinfo  {journal} {Class. Quant. Grav.}\ }\textbf {\bibinfo {volume}
  {25}},\ \bibinfo {pages} {065005} (\bibinfo {year} {2008})},\ \Eprint
  {https://arxiv.org/abs/0710.4369} {arXiv:0710.4369 [gr-qc]} \BibitemShut
  {NoStop}%
%%CITATION = ARXIV:0710.4369;%%
\bibitem [{\citenamefont {Vallisneri}\ and\ \citenamefont
  {Galley}(2012)}]{Vallisneri:2012np}%
  \BibitemOpen
  \bibfield  {author} {\bibinfo {author} {\bibfnamefont {M.}~\bibnamefont
  {Vallisneri}}\ and\ \bibinfo {author} {\bibfnamefont {C.~R.}\ \bibnamefont
  {Galley}},\ }\bibfield  {title} {\bibinfo {title} {{Non-sky-averaged
  sensitivity curves for space-based gravitational-wave observatories}},\
  }\href {https://doi.org/10.1088/0264-9381/29/12/124015} {\bibfield  {journal}
  {\bibinfo  {journal} {Class. Quant. Grav.}\ }\textbf {\bibinfo {volume}
  {29}},\ \bibinfo {pages} {124015} (\bibinfo {year} {2012})},\ \Eprint
  {https://arxiv.org/abs/1201.3684} {arXiv:1201.3684 [gr-qc]} \BibitemShut
  {NoStop}%
%%CITATION = ARXIV:1201.3684;%%
\bibitem [{\citenamefont {Tinto}\ and\ \citenamefont
  {da~Silva~Alves}(2010)}]{Tinto:2010hz}%
  \BibitemOpen
  \bibfield  {author} {\bibinfo {author} {\bibfnamefont {M.}~\bibnamefont
  {Tinto}}\ and\ \bibinfo {author} {\bibfnamefont {M.~E.}\ \bibnamefont
  {da~Silva~Alves}},\ }\bibfield  {title} {\bibinfo {title} {{LISA
  Sensitivities to Gravitational Waves from Relativistic Metric Theories of
  Gravity}},\ }\href {https://doi.org/10.1103/PhysRevD.82.122003} {\bibfield
  {journal} {\bibinfo  {journal} {Phys. Rev. D}\ }\textbf {\bibinfo {volume}
  {82}},\ \bibinfo {pages} {122003} (\bibinfo {year} {2010})},\ \Eprint
  {https://arxiv.org/abs/1010.1302} {arXiv:1010.1302 [gr-qc]} \BibitemShut
  {NoStop}%
\bibitem [{\citenamefont {Prince}\ \emph {et~al.}(2002)\citenamefont {Prince},
  \citenamefont {Tinto}, \citenamefont {Larson},\ and\ \citenamefont
  {Armstrong}}]{Prince:2002hp}%
  \BibitemOpen
  \bibfield  {author} {\bibinfo {author} {\bibfnamefont {T.~A.}\ \bibnamefont
  {Prince}}, \bibinfo {author} {\bibfnamefont {M.}~\bibnamefont {Tinto}},
  \bibinfo {author} {\bibfnamefont {S.~L.}\ \bibnamefont {Larson}},\ and\
  \bibinfo {author} {\bibfnamefont {J.~W.}\ \bibnamefont {Armstrong}},\
  }\bibfield  {title} {\bibinfo {title} {{The LISA optimal sensitivity}},\
  }\href {https://doi.org/10.1103/PhysRevD.66.122002} {\bibfield  {journal}
  {\bibinfo  {journal} {Phys. Rev. D}\ }\textbf {\bibinfo {volume} {66}},\
  \bibinfo {pages} {122002} (\bibinfo {year} {2002})},\ \Eprint
  {https://arxiv.org/abs/gr-qc/0209039} {arXiv:gr-qc/0209039 [gr-qc]}
  \BibitemShut {NoStop}%
%%CITATION = GR-QC/0209039;%%
\bibitem [{\citenamefont {Otto}\ \emph {et~al.}(2012)\citenamefont {Otto},
  \citenamefont {Heinzel},\ and\ \citenamefont {Danzmann}}]{Otto:2012dk}%
  \BibitemOpen
  \bibfield  {author} {\bibinfo {author} {\bibfnamefont {M.}~\bibnamefont
  {Otto}}, \bibinfo {author} {\bibfnamefont {G.}~\bibnamefont {Heinzel}},\ and\
  \bibinfo {author} {\bibfnamefont {K.}~\bibnamefont {Danzmann}},\ }\bibfield
  {title} {\bibinfo {title} {{TDI and clock noise removal for the split
  interferometry configuration of LISA}},\ }\href
  {https://doi.org/10.1088/0264-9381/29/20/205003} {\bibfield  {journal}
  {\bibinfo  {journal} {Class. Quant. Grav.}\ }\textbf {\bibinfo {volume}
  {29}},\ \bibinfo {pages} {205003} (\bibinfo {year} {2012})}\BibitemShut
  {NoStop}%
\bibitem [{\citenamefont {Otto}(2015)}]{Otto:2015}%
  \BibitemOpen
  \bibfield  {author} {\bibinfo {author} {\bibfnamefont {M.}~\bibnamefont
  {Otto}},\ }\emph {\bibinfo {title} {{Time-Delay Interferometry Simulations
  for the Laser Interferometer Space Antenna}}},\ \href
  {https://doi.org/10.15488/8545} {Ph.D. thesis},\ \bibinfo  {school} {Leibniz
  U., Hannover} (\bibinfo {year} {2015})\BibitemShut {NoStop}%
\bibitem [{\citenamefont {Tinto}\ and\ \citenamefont
  {Hartwig}(2018)}]{Tinto:2018kij}%
  \BibitemOpen
  \bibfield  {author} {\bibinfo {author} {\bibfnamefont {M.}~\bibnamefont
  {Tinto}}\ and\ \bibinfo {author} {\bibfnamefont {O.}~\bibnamefont
  {Hartwig}},\ }\bibfield  {title} {\bibinfo {title} {{Time-Delay
  Interferometry and Clock-Noise Calibration}},\ }\href
  {https://doi.org/10.1103/PhysRevD.98.042003} {\bibfield  {journal} {\bibinfo
  {journal} {Phys. Rev. D}\ }\textbf {\bibinfo {volume} {98}},\ \bibinfo
  {pages} {042003} (\bibinfo {year} {2018})},\ \Eprint
  {https://arxiv.org/abs/1807.02594} {arXiv:1807.02594 [gr-qc]} \BibitemShut
  {NoStop}%
\bibitem [{\citenamefont {Luo}\ \emph {et~al.}(2020)\citenamefont {Luo},
  \citenamefont {Guo}, \citenamefont {Jin}, \citenamefont {Wu},\ and\
  \citenamefont {Hu}}]{Luo:2020}%
  \BibitemOpen
  \bibfield  {author} {\bibinfo {author} {\bibfnamefont {Z.}~\bibnamefont
  {Luo}}, \bibinfo {author} {\bibfnamefont {Z.}~\bibnamefont {Guo}}, \bibinfo
  {author} {\bibfnamefont {G.}~\bibnamefont {Jin}}, \bibinfo {author}
  {\bibfnamefont {Y.}~\bibnamefont {Wu}},\ and\ \bibinfo {author}
  {\bibfnamefont {W.}~\bibnamefont {Hu}},\ }\bibfield  {title} {\bibinfo
  {title} {{A brief analysis to Taiji: Science and technology}},\ }\href
  {https://doi.org/doi.org/10.1016/j.rinp.2019.102918} {\bibfield  {journal}
  {\bibinfo  {journal} {Results in Physics}\ }\textbf {\bibinfo {volume}
  {16}},\ \bibinfo {pages} {102918} (\bibinfo {year} {2020})}\BibitemShut
  {NoStop}%
\bibitem [{\citenamefont {{Cutler}}\ and\ \citenamefont
  {{Flanagan}}(1994)}]{1994PhRvD..49.2658C}%
  \BibitemOpen
  \bibfield  {author} {\bibinfo {author} {\bibfnamefont {C.}~\bibnamefont
  {{Cutler}}}\ and\ \bibinfo {author} {\bibfnamefont {{\'E}.~E.}\ \bibnamefont
  {{Flanagan}}},\ }\bibfield  {title} {\bibinfo {title} {{Gravitational waves
  from merging compact binaries: How accurately can one extract the binary's
  parameters from the inspiral waveform?}},\ }\href
  {https://doi.org/10.1103/PhysRevD.49.2658} {\bibfield  {journal} {\bibinfo
  {journal} {\prd}\ }\textbf {\bibinfo {volume} {49}},\ \bibinfo {pages} {2658}
  (\bibinfo {year} {1994})},\ \Eprint {https://arxiv.org/abs/gr-qc/9402014}
  {arXiv:gr-qc/9402014 [gr-qc]} \BibitemShut {NoStop}%
\bibitem [{\citenamefont {Cutler}(1998)}]{Cutler:1997ta}%
  \BibitemOpen
  \bibfield  {author} {\bibinfo {author} {\bibfnamefont {C.}~\bibnamefont
  {Cutler}},\ }\bibfield  {title} {\bibinfo {title} {{Angular resolution of the
  LISA gravitational wave detector}},\ }\href
  {https://doi.org/10.1103/PhysRevD.57.7089} {\bibfield  {journal} {\bibinfo
  {journal} {Phys. Rev.}\ }\textbf {\bibinfo {volume} {D57}},\ \bibinfo {pages}
  {7089} (\bibinfo {year} {1998})},\ \Eprint
  {https://arxiv.org/abs/gr-qc/9703068} {arXiv:gr-qc/9703068 [gr-qc]}
  \BibitemShut {NoStop}%
%%CITATION = GR-QC/9703068;%%
\bibitem [{\citenamefont {Vallisneri}(2008)}]{Vallisneri:2007ev}%
  \BibitemOpen
  \bibfield  {author} {\bibinfo {author} {\bibfnamefont {M.}~\bibnamefont
  {Vallisneri}},\ }\bibfield  {title} {\bibinfo {title} {{Use and abuse of the
  Fisher information matrix in the assessment of gravitational-wave
  parameter-estimation prospects}},\ }\href
  {https://doi.org/10.1103/PhysRevD.77.042001} {\bibfield  {journal} {\bibinfo
  {journal} {Phys. Rev.}\ }\textbf {\bibinfo {volume} {D77}},\ \bibinfo {pages}
  {042001} (\bibinfo {year} {2008})},\ \Eprint
  {https://arxiv.org/abs/gr-qc/0703086} {arXiv:gr-qc/0703086 [GR-QC]}
  \BibitemShut {NoStop}%
%%CITATION = GR-QC/0703086;%%
\bibitem [{\citenamefont {Kuns}\ \emph {et~al.}(2019)\citenamefont {Kuns},
  \citenamefont {Yu}, \citenamefont {Chen},\ and\ \citenamefont
  {Adhikari}}]{Kuns:2019upi}%
  \BibitemOpen
  \bibfield  {author} {\bibinfo {author} {\bibfnamefont {K.~A.}\ \bibnamefont
  {Kuns}}, \bibinfo {author} {\bibfnamefont {H.}~\bibnamefont {Yu}}, \bibinfo
  {author} {\bibfnamefont {Y.}~\bibnamefont {Chen}},\ and\ \bibinfo {author}
  {\bibfnamefont {R.~X.}\ \bibnamefont {Adhikari}},\ }\bibfield  {title}
  {\bibinfo {title} {{Astrophysics and cosmology with a deci-hertz
  gravitational-wave detector: TianGO}},\ }\href@noop {} {\  (\bibinfo {year}
  {2019})},\ \Eprint {https://arxiv.org/abs/1908.06004} {arXiv:1908.06004
  [gr-qc]} \BibitemShut {NoStop}%
%%CITATION = ARXIV:1908.06004;%%
\bibitem [{\citenamefont {{Khan}}\ \emph {et~al.}(2016)\citenamefont {{Khan}},
  \citenamefont {{Husa}}, \citenamefont {{Hannam}}, \citenamefont {{Ohme}},
  \citenamefont {{P{\"u}rrer}}, \citenamefont {{Forteza}},\ and\ \citenamefont
  {{Boh{\'e}}}}]{Khan:2015jqa}%
  \BibitemOpen
  \bibfield  {author} {\bibinfo {author} {\bibfnamefont {S.}~\bibnamefont
  {{Khan}}}, \bibinfo {author} {\bibfnamefont {S.}~\bibnamefont {{Husa}}},
  \bibinfo {author} {\bibfnamefont {M.}~\bibnamefont {{Hannam}}}, \bibinfo
  {author} {\bibfnamefont {F.}~\bibnamefont {{Ohme}}}, \bibinfo {author}
  {\bibfnamefont {M.}~\bibnamefont {{P{\"u}rrer}}}, \bibinfo {author}
  {\bibfnamefont {X.~J.}\ \bibnamefont {{Forteza}}},\ and\ \bibinfo {author}
  {\bibfnamefont {A.}~\bibnamefont {{Boh{\'e}}}},\ }\bibfield  {title}
  {\bibinfo {title} {{Frequency-domain gravitational waves from nonprecessing
  black-hole binaries. II. A phenomenological model for the advanced detector
  era}},\ }\href {https://doi.org/10.1103/PhysRevD.93.044007} {\bibfield
  {journal} {\bibinfo  {journal} {\prd}\ }\textbf {\bibinfo {volume} {93}},\
  \bibinfo {eid} {044007} (\bibinfo {year} {2016})},\ \Eprint
  {https://arxiv.org/abs/1508.07253} {arXiv:1508.07253 [gr-qc]} \BibitemShut
  {NoStop}%
\bibitem [{\citenamefont {{Chatziioannou}}\ \emph {et~al.}(2012)\citenamefont
  {{Chatziioannou}}, \citenamefont {{Yunes}},\ and\ \citenamefont
  {{Cornish}}}]{yunes2012PhRvD..86b2004C}%
  \BibitemOpen
  \bibfield  {author} {\bibinfo {author} {\bibfnamefont {K.}~\bibnamefont
  {{Chatziioannou}}}, \bibinfo {author} {\bibfnamefont {N.}~\bibnamefont
  {{Yunes}}},\ and\ \bibinfo {author} {\bibfnamefont {N.}~\bibnamefont
  {{Cornish}}},\ }\bibfield  {title} {\bibinfo {title} {{Model-independent test
  of general relativity: An extended post-Einsteinian framework with complete
  polarization content}},\ }\href {https://doi.org/10.1103/PhysRevD.86.022004}
  {\bibfield  {journal} {\bibinfo  {journal} {\prd}\ }\textbf {\bibinfo
  {volume} {86}},\ \bibinfo {eid} {022004} (\bibinfo {year} {2012})},\ \Eprint
  {https://arxiv.org/abs/1204.2585} {arXiv:1204.2585 [gr-qc]} \BibitemShut
  {NoStop}%
\bibitem [{\citenamefont {Chatziioannou}\ \emph {et~al.}(2017)\citenamefont
  {Chatziioannou}, \citenamefont {Yunes},\ and\ \citenamefont
  {Cornish}}]{PhysRevD.95.129901}%
  \BibitemOpen
  \bibfield  {author} {\bibinfo {author} {\bibfnamefont {K.}~\bibnamefont
  {Chatziioannou}}, \bibinfo {author} {\bibfnamefont {N.}~\bibnamefont
  {Yunes}},\ and\ \bibinfo {author} {\bibfnamefont {N.}~\bibnamefont
  {Cornish}},\ }\bibfield  {title} {\bibinfo {title} {Erratum:
  Model-independent test of general relativity: An extended post-einsteinian
  framework with complete polarization content [phys. rev. d 86, 022004
  (2012)]},\ }\href {https://doi.org/10.1103/PhysRevD.95.129901} {\bibfield
  {journal} {\bibinfo  {journal} {Phys. Rev. D}\ }\textbf {\bibinfo {volume}
  {95}},\ \bibinfo {pages} {129901} (\bibinfo {year} {2017})}\BibitemShut
  {NoStop}%
\bibitem [{\citenamefont {Will}(1998)}]{Will:1997bb}%
  \BibitemOpen
  \bibfield  {author} {\bibinfo {author} {\bibfnamefont {C.~M.}\ \bibnamefont
  {Will}},\ }\bibfield  {title} {\bibinfo {title} {{Bounding the mass of the
  graviton using gravitational wave observations of inspiralling compact
  binaries}},\ }\href {https://doi.org/10.1103/PhysRevD.57.2061} {\bibfield
  {journal} {\bibinfo  {journal} {Phys. Rev. D}\ }\textbf {\bibinfo {volume}
  {57}},\ \bibinfo {pages} {2061} (\bibinfo {year} {1998})},\ \Eprint
  {https://arxiv.org/abs/gr-qc/9709011} {arXiv:gr-qc/9709011} \BibitemShut
  {NoStop}%
\bibitem [{\citenamefont {Will}\ and\ \citenamefont
  {Yunes}(2004)}]{Will:2004xi}%
  \BibitemOpen
  \bibfield  {author} {\bibinfo {author} {\bibfnamefont {C.~M.}\ \bibnamefont
  {Will}}\ and\ \bibinfo {author} {\bibfnamefont {N.}~\bibnamefont {Yunes}},\
  }\bibfield  {title} {\bibinfo {title} {{Testing alternative theories of
  gravity using LISA}},\ }\href {https://doi.org/10.1088/0264-9381/21/18/006}
  {\bibfield  {journal} {\bibinfo  {journal} {Class. Quant. Grav.}\ }\textbf
  {\bibinfo {volume} {21}},\ \bibinfo {pages} {4367} (\bibinfo {year}
  {2004})},\ \Eprint {https://arxiv.org/abs/gr-qc/0403100}
  {arXiv:gr-qc/0403100} \BibitemShut {NoStop}%
\bibitem [{\citenamefont {Berti}\ \emph {et~al.}(2005)\citenamefont {Berti},
  \citenamefont {Buonanno},\ and\ \citenamefont {Will}}]{Berti:2005qd}%
  \BibitemOpen
  \bibfield  {author} {\bibinfo {author} {\bibfnamefont {E.}~\bibnamefont
  {Berti}}, \bibinfo {author} {\bibfnamefont {A.}~\bibnamefont {Buonanno}},\
  and\ \bibinfo {author} {\bibfnamefont {C.~M.}\ \bibnamefont {Will}},\
  }\bibfield  {title} {\bibinfo {title} {{Testing general relativity and
  probing the merger history of massive black holes with LISA}},\ }\href
  {https://doi.org/10.1088/0264-9381/22/18/S08} {\bibfield  {journal} {\bibinfo
   {journal} {Class. Quant. Grav.}\ }\textbf {\bibinfo {volume} {22}},\
  \bibinfo {pages} {S943} (\bibinfo {year} {2005})},\ \Eprint
  {https://arxiv.org/abs/gr-qc/0504017} {arXiv:gr-qc/0504017} \BibitemShut
  {NoStop}%
\bibitem [{\citenamefont {Stavridis}\ and\ \citenamefont
  {Will}(2009)}]{Stavridis:2009mb}%
  \BibitemOpen
  \bibfield  {author} {\bibinfo {author} {\bibfnamefont {A.}~\bibnamefont
  {Stavridis}}\ and\ \bibinfo {author} {\bibfnamefont {C.~M.}\ \bibnamefont
  {Will}},\ }\bibfield  {title} {\bibinfo {title} {{Bounding the mass of the
  graviton with gravitational waves: Effect of spin precessions in massive
  black hole binaries}},\ }\href {https://doi.org/10.1103/PhysRevD.80.044002}
  {\bibfield  {journal} {\bibinfo  {journal} {Phys. Rev. D}\ }\textbf {\bibinfo
  {volume} {80}},\ \bibinfo {pages} {044002} (\bibinfo {year} {2009})},\
  \Eprint {https://arxiv.org/abs/0906.3602} {arXiv:0906.3602 [gr-qc]}
  \BibitemShut {NoStop}%
\bibitem [{\citenamefont {Arun}\ and\ \citenamefont
  {Will}(2009)}]{Arun:2009pq}%
  \BibitemOpen
  \bibfield  {author} {\bibinfo {author} {\bibfnamefont {K.~G.}\ \bibnamefont
  {Arun}}\ and\ \bibinfo {author} {\bibfnamefont {C.~M.}\ \bibnamefont
  {Will}},\ }\bibfield  {title} {\bibinfo {title} {{Bounding the mass of the
  graviton with gravitational waves: Effect of higher harmonics in
  gravitational waveform templates}},\ }\href
  {https://doi.org/10.1088/0264-9381/26/15/155002} {\bibfield  {journal}
  {\bibinfo  {journal} {Class. Quant. Grav.}\ }\textbf {\bibinfo {volume}
  {26}},\ \bibinfo {pages} {155002} (\bibinfo {year} {2009})},\ \Eprint
  {https://arxiv.org/abs/0904.1190} {arXiv:0904.1190 [gr-qc]} \BibitemShut
  {NoStop}%
\bibitem [{\citenamefont {Keppel}\ and\ \citenamefont
  {Ajith}(2010)}]{Keppel:2010qu}%
  \BibitemOpen
  \bibfield  {author} {\bibinfo {author} {\bibfnamefont {D.}~\bibnamefont
  {Keppel}}\ and\ \bibinfo {author} {\bibfnamefont {P.}~\bibnamefont {Ajith}},\
  }\bibfield  {title} {\bibinfo {title} {{Constraining the mass of the graviton
  using coalescing black-hole binaries}},\ }\href
  {https://doi.org/10.1103/PhysRevD.82.122001} {\bibfield  {journal} {\bibinfo
  {journal} {Phys. Rev. D}\ }\textbf {\bibinfo {volume} {82}},\ \bibinfo
  {pages} {122001} (\bibinfo {year} {2010})},\ \Eprint
  {https://arxiv.org/abs/1004.0284} {arXiv:1004.0284 [gr-qc]} \BibitemShut
  {NoStop}%
\bibitem [{\citenamefont {Yagi}\ and\ \citenamefont
  {Tanaka}(2010)}]{Yagi:2009zm}%
  \BibitemOpen
  \bibfield  {author} {\bibinfo {author} {\bibfnamefont {K.}~\bibnamefont
  {Yagi}}\ and\ \bibinfo {author} {\bibfnamefont {T.}~\bibnamefont {Tanaka}},\
  }\bibfield  {title} {\bibinfo {title} {{Constraining alternative theories of
  gravity by gravitational waves from precessing eccentric compact binaries
  with LISA}},\ }\href {https://doi.org/10.1103/PhysRevD.81.109902} {\bibfield
  {journal} {\bibinfo  {journal} {Phys. Rev. D}\ }\textbf {\bibinfo {volume}
  {81}},\ \bibinfo {pages} {064008} (\bibinfo {year} {2010})},\ \bibinfo {note}
  {[Erratum: Phys.Rev.D 81, 109902 (2010)]},\ \Eprint
  {https://arxiv.org/abs/0906.4269} {arXiv:0906.4269 [gr-qc]} \BibitemShut
  {NoStop}%
\bibitem [{\citenamefont {Cornish}\ \emph {et~al.}(2011)\citenamefont
  {Cornish}, \citenamefont {Sampson}, \citenamefont {Yunes},\ and\
  \citenamefont {Pretorius}}]{Cornish:2011ys}%
  \BibitemOpen
  \bibfield  {author} {\bibinfo {author} {\bibfnamefont {N.}~\bibnamefont
  {Cornish}}, \bibinfo {author} {\bibfnamefont {L.}~\bibnamefont {Sampson}},
  \bibinfo {author} {\bibfnamefont {N.}~\bibnamefont {Yunes}},\ and\ \bibinfo
  {author} {\bibfnamefont {F.}~\bibnamefont {Pretorius}},\ }\bibfield  {title}
  {\bibinfo {title} {{Gravitational Wave Tests of General Relativity with the
  Parameterized Post-Einsteinian Framework}},\ }\href
  {https://doi.org/10.1103/PhysRevD.84.062003} {\bibfield  {journal} {\bibinfo
  {journal} {Phys. Rev. D}\ }\textbf {\bibinfo {volume} {84}},\ \bibinfo
  {pages} {062003} (\bibinfo {year} {2011})},\ \Eprint
  {https://arxiv.org/abs/1105.2088} {arXiv:1105.2088 [gr-qc]} \BibitemShut
  {NoStop}%
\bibitem [{\citenamefont {Flanagan}(1993)}]{Flanagan:1993ix}%
  \BibitemOpen
  \bibfield  {author} {\bibinfo {author} {\bibfnamefont {E.~E.}\ \bibnamefont
  {Flanagan}},\ }\bibfield  {title} {\bibinfo {title} {{The Sensitivity of the
  laser interferometer gravitational wave observatory (LIGO) to a stochastic
  background, and its dependence on the detector orientations}},\ }\href
  {https://doi.org/10.1103/PhysRevD.48.2389} {\bibfield  {journal} {\bibinfo
  {journal} {Phys. Rev. D}\ }\textbf {\bibinfo {volume} {48}},\ \bibinfo
  {pages} {2389} (\bibinfo {year} {1993})},\ \Eprint
  {https://arxiv.org/abs/astro-ph/9305029} {arXiv:astro-ph/9305029}
  \BibitemShut {NoStop}%
\bibitem [{\citenamefont {Christensen}(1992)}]{Christensen:1992wi}%
  \BibitemOpen
  \bibfield  {author} {\bibinfo {author} {\bibfnamefont {N.}~\bibnamefont
  {Christensen}},\ }\bibfield  {title} {\bibinfo {title} {{Measuring the
  stochastic gravitational radiation background with laser interferometric
  antennas}},\ }\href {https://doi.org/10.1103/PhysRevD.46.5250} {\bibfield
  {journal} {\bibinfo  {journal} {Phys. Rev. D}\ }\textbf {\bibinfo {volume}
  {46}},\ \bibinfo {pages} {5250} (\bibinfo {year} {1992})}\BibitemShut
  {NoStop}%
\bibitem [{\citenamefont {Whelan}\ \emph {et~al.}(2002)\citenamefont {Whelan},
  \citenamefont {Anderson}, \citenamefont {Casquette}, \citenamefont {Diaz},
  \citenamefont {Heng}, \citenamefont {McHugh}, \citenamefont {Romano},
  \citenamefont {Torres Charlie~W.}, \citenamefont {Trejo},\ and\ \citenamefont
  {Vecchio}}]{Whelan:2001qw}%
  \BibitemOpen
  \bibfield  {author} {\bibinfo {author} {\bibfnamefont {J.~T.}\ \bibnamefont
  {Whelan}}, \bibinfo {author} {\bibfnamefont {W.~G.}\ \bibnamefont
  {Anderson}}, \bibinfo {author} {\bibfnamefont {M.}~\bibnamefont {Casquette}},
  \bibinfo {author} {\bibfnamefont {M.~C.}\ \bibnamefont {Diaz}}, \bibinfo
  {author} {\bibfnamefont {I.~S.}\ \bibnamefont {Heng}}, \bibinfo {author}
  {\bibfnamefont {M.}~\bibnamefont {McHugh}}, \bibinfo {author} {\bibfnamefont
  {J.~D.}\ \bibnamefont {Romano}}, \bibinfo {author} {\bibfnamefont
  {J.}~\bibnamefont {Torres Charlie~W.}}, \bibinfo {author} {\bibfnamefont
  {R.~M.}\ \bibnamefont {Trejo}},\ and\ \bibinfo {author} {\bibfnamefont
  {A.}~\bibnamefont {Vecchio}},\ }\bibfield  {title} {\bibinfo {title}
  {{Progress on stochastic background search codes for LIGO}},\ }\href
  {https://doi.org/10.1088/0264-9381/19/7/339} {\bibfield  {journal} {\bibinfo
  {journal} {Class. Quant. Grav.}\ }\textbf {\bibinfo {volume} {19}},\ \bibinfo
  {pages} {1521} (\bibinfo {year} {2002})},\ \Eprint
  {https://arxiv.org/abs/gr-qc/0110019} {arXiv:gr-qc/0110019} \BibitemShut
  {NoStop}%
\bibitem [{\citenamefont {Adams}\ and\ \citenamefont
  {Cornish}(2010)}]{Adams:2010vc}%
  \BibitemOpen
  \bibfield  {author} {\bibinfo {author} {\bibfnamefont {M.~R.}\ \bibnamefont
  {Adams}}\ and\ \bibinfo {author} {\bibfnamefont {N.~J.}\ \bibnamefont
  {Cornish}},\ }\bibfield  {title} {\bibinfo {title} {{Discriminating between a
  Stochastic Gravitational Wave Background and Instrument Noise}},\ }\href
  {https://doi.org/10.1103/PhysRevD.82.022002} {\bibfield  {journal} {\bibinfo
  {journal} {Phys. Rev. D}\ }\textbf {\bibinfo {volume} {82}},\ \bibinfo
  {pages} {022002} (\bibinfo {year} {2010})},\ \Eprint
  {https://arxiv.org/abs/1002.1291} {arXiv:1002.1291 [gr-qc]} \BibitemShut
  {NoStop}%
\bibitem [{\citenamefont {Ni}(2013)}]{Ni:2012eh}%
  \BibitemOpen
  \bibfield  {author} {\bibinfo {author} {\bibfnamefont {W.-T.}\ \bibnamefont
  {Ni}},\ }\bibfield  {title} {\bibinfo {title} {{ASTROD-GW: Overview and
  Progress}},\ }\href {https://doi.org/10.1142/S0218271813410046} {\bibfield
  {journal} {\bibinfo  {journal} {Int. J. Mod. Phys. D}\ }\textbf {\bibinfo
  {volume} {22}},\ \bibinfo {pages} {1341004} (\bibinfo {year} {2013})},\
  \Eprint {https://arxiv.org/abs/1212.2816} {arXiv:1212.2816 [astro-ph.IM]}
  \BibitemShut {NoStop}%
\bibitem [{\citenamefont {Baker}\ \emph {et~al.}(2019)\citenamefont {Baker}
  \emph {et~al.}}]{Baker:2019pnp}%
  \BibitemOpen
  \bibfield  {author} {\bibinfo {author} {\bibfnamefont {J.}~\bibnamefont
  {Baker}} \emph {et~al.},\ }\bibfield  {title} {\bibinfo {title} {{Space Based
  Gravitational Wave Astronomy Beyond LISA}},\ }\href@noop {} {\  (\bibinfo
  {year} {2019})},\ \Eprint {https://arxiv.org/abs/1907.11305}
  {arXiv:1907.11305 [astro-ph.IM]} \BibitemShut {NoStop}%
\bibitem [{\citenamefont {Sesana}\ \emph {et~al.}(2019)\citenamefont {Sesana}
  \emph {et~al.}}]{Sesana:2019vho}%
  \BibitemOpen
  \bibfield  {author} {\bibinfo {author} {\bibfnamefont {A.}~\bibnamefont
  {Sesana}} \emph {et~al.},\ }\bibfield  {title} {\bibinfo {title} {{Unveiling
  the Gravitational Universe at \textbackslash{}mu-Hz Frequencies}},\
  }\href@noop {} {\  (\bibinfo {year} {2019})},\ \Eprint
  {https://arxiv.org/abs/1908.11391} {arXiv:1908.11391 [astro-ph.IM]}
  \BibitemShut {NoStop}%
\end{thebibliography}%
%\bibliography{apsempty}% Produces the bibliography via BibTeX.

\end{document}